\documentclass[fleqn,usenatbib]{rasti}
\usepackage{booktabs}
\usepackage{multirow}

\usepackage{newtxtext,newtxmath}
\usepackage[T1]{fontenc}

\DeclareRobustCommand{\VAN}[3]{#2}
\let\VANthebibliography\thebibliography
\def\thebibliography{\DeclareRobustCommand{\VAN}[3]{##3}\VANthebibliography}

\usepackage{graphicx}	% Including figure files
\usepackage{amsmath}	% Advanced maths commands
\usepackage{aas_macros}

\title[Tied-Array Beam Localisation]{Tied-Array Beam Localisation of Radio Transients and Pulsars}

\author[M. C. Bezuidenhout et al.]{M.~C.~Bezuidenhout,$^{1,2}$\thanks{E-mail: mechiel.bez@gmail.com}
C.~J.~Clark$^{3,4,2}$,
R.~.P.~Breton$^{2}$,
B.~W.~Stappers$^{2}$,
E.~D.~Barr$^{5}$,
M.~Caleb$^{6}$,
W.~Chen$^{5}$,
\newauthor
F.~Jankowski$^{2,7}$,
M.~Kramer$^{3}$,
K.~Rajwade$^{8}$, and
M.~Surnis$^{2,9}$
\\
$^{1}$Centre for Space Research, North-West University, Potchefstroom 2351, South Africa\\
$^{2}$Jodrell Bank Centre for Astrophysics, University of Manchester, Manchester M13 9PL, UK\\
$^{3}$Max Planck Institute for Gravitational Physics (Albert Einstein Institute), D-30167 Hannover, Germany\\
$^{4}$Leibniz Universit\"at Hannover, D-30167 Hannover, Germany\\
$^{5}$Max-Planck-Institut f\"ur Radioastronomie, Auf dem Hugel 69, D-53121 Bonn, Germany\\
$^{6}$Sydney Institute for Astronomy, School of Physics, The University of Sydney, NSW 2006, Australia\\
$^{7}$LPC2E, Universit\'{e} d'Orl\'{e}ans, CNRS, 3A Avenue de la Recherche Scientifique, 45071 Orl\'{e}ans, France\\
$^{8}$ASTRON, the Netherlands Institute for Radio Astronomy, Oude Hoogeveensedijk 4, 7991 PD Dwingeloo, The Netherlands\\
$^{9}$Department of Physics, IISER Bhopal, Bhauri Bypass Road, Bhopal, 462066, India
}

\date{Accepted 13.02.2023. Received 27.01.2023; in original form 20.05.2022}

\pubyear{2022}

\begin{document}
\label{firstpage}
\pagerange{\pageref{firstpage}--\pageref{lastpage}}
\maketitle

\begin{abstract}
Multi-element interferometers such as MeerKAT, which observe with high time resolution and have a wide field-of-view, provide an ideal opportunity to perform real-time, untargeted transient and pulsar searches. However, because of data storage limitations, it is not always feasible to store the baseband data required to image the field of a discovered transient or pulsar. This limits the ability of surveys to effectively localise their discoveries and may restrict opportunities for follow-up science, especially of one-off events like some Fast Radio Bursts (FRBs). Here we present a novel maximum-likelihood estimation approach to localising transients and pulsars detected in multiple MeerKAT tied-array beams at once, which we call Tied Array Beam Localisation (TABLo), as well as a Python implementation of the method named \textsc{SeeKAT}. We provide real-world examples of \textsc{SeeKAT}'s use as well as a Monte Carlo analysis to show that it is capable of localising single pulses detected in beamformed MeerKAT data to (sub-)arcsecond precision.   
\end{abstract}


%%%%%%%%%%%%%%%%%%%%%%%%%%%%%%%%%%%%%%%%%%%%%%%%%%

%%%%%%%%%%%%%%%%% BODY OF PAPER %%%%%%%%%%%%%%%%%%

%-----------------------------------------------------------SECTION 1---------------------------------------------------------------------------------------------------------------

\section{Introduction}
\label{sec:Intro}

Starting with the discovery of the first radio pulsar in 1968 \citep{1968hewish}, the advent of high time resolution astronomical observations has allowed for the study of a broad class of astrophysical objects that are now collectively referred to as radio transients. A subset of radio transients, whose emission varies on time scales of seconds or less, are classified as \textit{fast} radio transients \citep[][]{cordes2007}. The short light travel times suggested by this rapid variability point to very small emission regions.
This, combined with the vast amounts of energy released, naturally leads to fast transients being associated with the extreme environments of compact objects like neutron stars and black holes.

One-off or irregularly repeating fast transients are most easily observed in single pulse searches. These sources include Rotating RAdio Transients (RRATs), magnetars, and extragalactic Fast Radio Bursts (FRBs). The degree to which fast transient single pulses are dispersed by their frequency-dependent propagation through the intervening matter, given by their dispersion measures (DMs), can serve as a probe of the extragalactic medium. For example, DM values of distant FRBs can be used in concert with independent distance measurements to estimate cosmic baryon densities \citep{2020macquart}. \citet{2019walters} showed that independent measurements of the redshift of $\sim$100 FRB host galaxies are required to make a measurement of the diffuse gas fraction in the Galaxy, which would help alleviate the ``missing baryon problem'' in the current $\Lambda$CDM model of the universe. DM measurements of FRBs with host galaxy associations have also been used to constrain the properties of intervening galactic halo gas \citep[][]{2019prochaska}. To obtain independent distance measurements, FRBs must be unambiguously associated with host galaxies, which requires localisation to a precision of $\sim$1~arcsecond \citep[][]{2017eftekhari}.

Precise localisation can also enable improved follow-up observations of pulsars and repeating fast transients with more accurate source targeting that maximises the telescope gain at the source position, leading to higher measured signal-to-noise (S/N) ratios. Additionally, localisation can help resolve the problem of the covariance of the pulse period derivative ($\Dot{P}$) and source position. This can enable more robust $\Dot{P}$ measurements in cases where observing time for follow-up timing is scarce, as well as for sparsely-observed periodic sources like RRATs and radio-loud magnetars. 

Typically, transients and pulsars can be localised through radio synthesis imaging using complex voltage data captured from the array elements \citep[see e.g.][]{1984pearson,2019bannister}. Untargeted fast transient surveys, which make use of tied-array beamforming to perform single pulse searches (often in real time), may employ transient buffers that store snippets of imaging data around candidate pulse times for the purposes of off-line localisation. However, the memory demands of recording imaging data often becomes prohibitive, especially for surveys that form hundreds of tied-array beams (TABs) and search up to very high DMs. Such surveys must rely on alternative methods for localising discovered transients.

It is also helpful to be able to localise a newly-discovered transient or pulsar quickly with sufficient precision to trigger rapid follow-up observations at higher frequencies or with instruments requiring precise positional information. Rapid follow-up observations may reveal repeat pulses, multi-wavelength counterparts, and fast-fading afterglows \citep[][]{2014yi,2016vedantham}.

It is therefore valuable to have a method for localising transients observed in beamformed data to a precision smaller than the size of the half-power beam width (HPBW) of a TAB. The first instance of such a method was used by the Westerbork Synthesis Radio Telescope (WSRT) \textsc{8gr8} survey, where they localised detected sources to the crossing point of fan-beams at different hour angles \citep[][]{2013rubioherrera}. Subsequently, \citet{2014spitler} used spatial information about the gain of the Arecibo ALFA receiver at different frequencies to map out the instrument's spectral index, and based on the observed spectral index of an FRB~20121102A detection determined that it must have occurred on the rising edge of the first sidelobe.

\citet{2015obrocka} and \citet[][]{2015obrockaThesis}\footnote{PhD thesis available online at \url{https://ethos.bl.uk/OrderDetails.do?uin=uk.bl.ethos.647424}} expanded on this idea by applying it to multibeam detections with interferometric arrays. This method stipulates that, for a given pair of TABs, possible source positions are defined as those where both the observed ratio of S/N values and difference of spectral indices match the expected values derived from modelling of the TABs' spatially-dependent and frequency-dependent gain within 1-$\sigma$ errors. The final localisation is then taken as the intersection of the localisation contours derived for each pair of TABs in which a pulse was detected above the detection threshold. \citet[][]{2015obrocka} demonstrated the ability of this method to localise a single pulse observed by the MeerKAT telescope to $\sim$arcsecond precision. \citet[][]{2019petroff} used an adapted version of this method to pinpoint FRB 20110214A detected by the Parkes telescope to one of three $\sim$10 arcminute$^2$ regions.

A similar method was used to constrain a single pulse from FRB 20170107A detected by the Australian Square Kilometre Array Precursor (ASKAP) to a region 8~arcminutes across with 90~per~cent confidence \citep[][]{2017bannister}. In this case, the authors assumed a simple Gaussian beam shape to model the expected flux density of the pulse over space, and compared this to the measured flux density in the primary beam and eight adjacent TABs. They applied a Bayesian methodology to sample the posterior positional probability. \citet[][]{2018shannon} and \citet[][]{2019qiu} have used the same method to localise several ASKAP-detected FRBs to $\sim$arcminute to $\sim$10~arcminute precision.

% Probabilistic localisation of multiple-beam single-pulse detections based on the distribution of observed flux densities has already been demonstrated for the ASKAP \citep[][]{2017bannister} and CHIME \citep[][]{2021chimecat} telescopes. In both cases, the FRBs could be localised to within a region smaller than the primary beam size ($\sim$arcminute precision), but not to within the $\sim1$~arcsecond precision required to associate an FRB with a host galaxy at redshift $z=0.5$ with a chance coincidence probability of $<1$~per~cent \citep[][]{2020macquart}. In this work we expand on a similar probabilistic localisation method first proposed for MeerKAT detections by \citet{2015obrocka}. We have developed an implementation of this method that is capable of localising multibeam MeerKAT detections to arcsecond or subarcsecond precision.

The Canadian Hydrogen Intensity Mapping Experiment (CHIME) have also used a variation of this method to constrain some of their FRB discoveries to polygonal regions of several arcminutes to degrees  \citep[][]{2019ChimeRepeater,2019Chime8Repeaters, 2021michilli}. They perform post-facto re-phasing of the received radio waves to various trial positions surrounding a detected pulse, and measure one S/N value for each formed beam. The beams are modelled using 2D sinc$^{2}$ functions, and the expected response ratios are fit to the derived S/N values using $\chi^{2}$ minimisation. Confidence regions are taken corresponding to contours of equal $\chi^{2}$ value.

In this paper, we present a new localisation method, named Tied-Array Beam Localisation (TABLo), developed to rapidly localise sources discovered by the MeerTRAP real-time fast transient and pulsar search using MeerKAT \citep[Stappers et al.\ in prep.;][]{2018sanidas,2022bezuidenhout}, as well as the TRAnsients and PUlsars with MeerKAT (TRAPUM) survey \citep[][]{2016stappers}. TABLo combines the \citet[][]{2015obrocka} approach of utilising the distribution of S/N values of a detected pulse in adjacent beams with a maximum likelihood estimation (MLE) approach similar to that of the CHIME collaboration. This method eschews the spectral index difference fitting of \citet[][]{2015obrocka} for reasons explored in $\S$\ref{sec:spec_ind}, as well as the phase-referencing grid approach described in \citet[][]{2021michilli} for CHIME FRBs. We show that we can obtain (sub-)arcsecond-level localisations with robust uncertainty estimation quickly enough for real-time use and rapid follow-up triggering.

$\S\ref{sec:beam}$ illustrates how the spatially dependent gain of TABs can be accurately modelled, $\S\ref{sec:tablo}$ shows how TABLo uses those beam models for localisation and describes a software implementation of the method, and $\S\ref{sec:results}$ presents selected results from its application to MeerKAT observations. Finally, $\S\ref{sec:elaborations}$ outlines possible future developments of this method working towards improved localisation precision.

% -----------------------------------------------------------SECTION 2---------------------------------------------------------------------------------------------------------------

\section{MeerKAT beamforming}\label{sec:beam}

\begin{figure*}
    \centering
    \includegraphics[width=\textwidth]{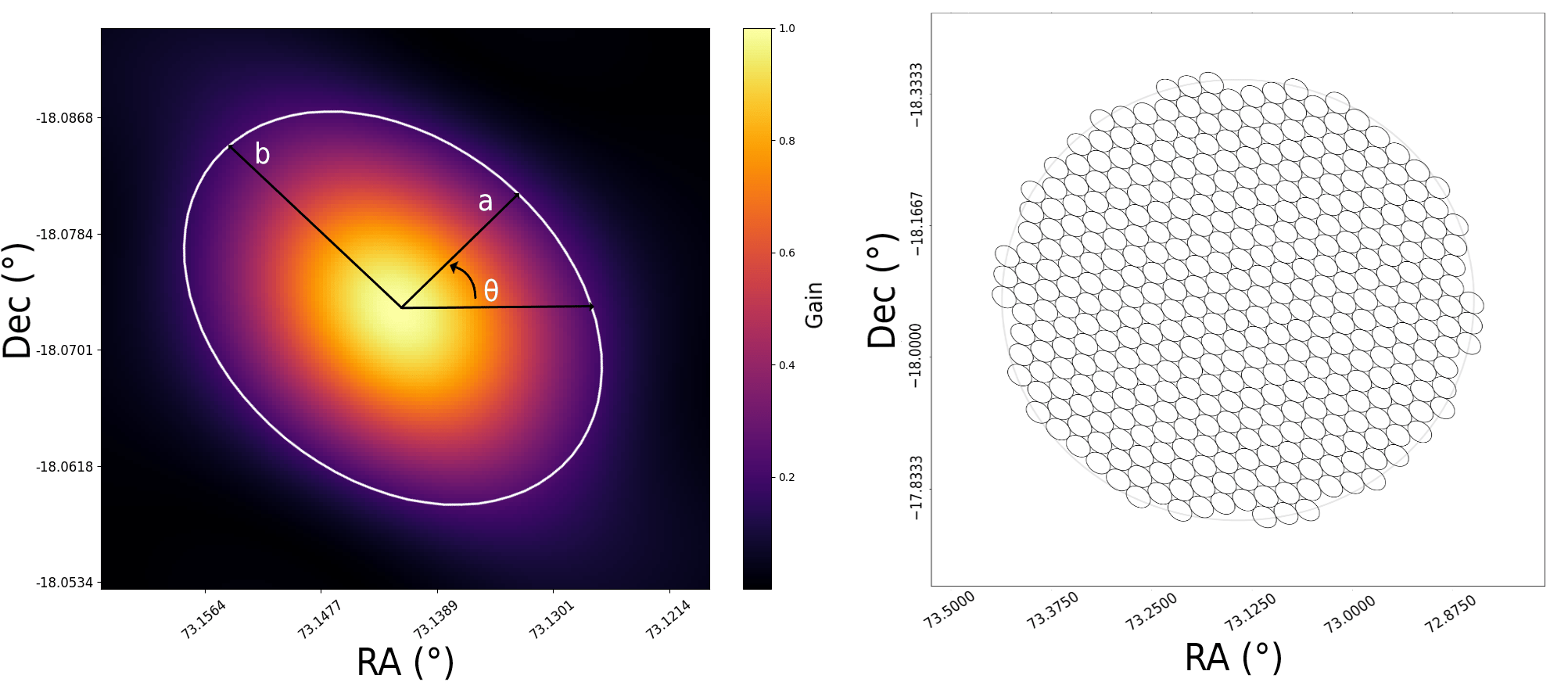}
    \caption{Images from a beamforming simulation performed using the \textsc{Mosaic} package for a MeerKAT observation at 11:30:52 UTC on 2020-07-16. During the observation, 384 TABs were tiled centred at RA 04:52:34.11 and Dec $-$18:04:23.4 to overlap at their 25~per~cent gain level at 1.284~GHz. \textbf{Left:} The instantaneous simulated PSF of one TAB at 1.284~GHz. The colour scale indicates the relative gain of the TAB. The white contour is an ellipse fit to the 25 per cent level of the PSF, with its semi-minor, semi-major, and position angle indicated by $a$, $b$, and $\theta$, respectively. \textbf{Right:} The tiling pattern of TABs within the primary beam, where the inner ellipses represent the 25 per cent response level of the TABs, all of which are assumed to be identical. The outer grey circle indicates the edge of the tiling pattern (not the edge of the PSFs). In this case, the TAB tiling pattern extends to approximately the 55~per~cent gain level of the MeerKAT primary beam at 1.284~GHz.}
    \label{fig:tilesim}
\end{figure*}

For parabolic-dish-based interferometers like MeerKAT and ASKAP, each steerable dish collects electromagnetic signals from a portion of the sky contained within the so-called primary beam, the extent of which is determined by the dish size. If a dish of diameter $D$ observes signals with a wavelength $\lambda$, then its primary beam's relative sensitivity to those signals will take the form of an Airy disk function over the sky, with the main lobe having a half-power point given, to first order, by $\theta_{\mathrm{PB}}\sim\lambda/D$ in radians \citep[see e.g.][]{2014burke}. The signals from all elements of the array can be combined coherently to form an aperture array. By introducing delays to each element's response before weighted addition, TABs can be formed with sizes determined instead by the maximum baseline\footnote{In practice, the actual size of the beam also depends on the distribution of the dishes and the weighting scheme used to combine their signals.} $d$, so that $\theta_{\mathrm{TAB}}\sim\lambda/d$. Since $d \gg D$, $\theta_{\mathrm{TAB}} \ll \theta_{\mathrm{PB}}$, thus achieving much finer angular resolution. Through controlling the delays added to each dish's signal, hundreds of TABs can be formed at different positions within the primary beam at once, thus attaining a collective FoV comparable to that of an individual dish with a much better angular resolution and sensitivity  \citep[see e.g.][]{2021chen}.

% \textcolor{blue}{
Beamforming observations with MeerKAT can be performed using the Filterbank and BeamForming User-Supplied Equipment (FBFUSE) compute cluster developed by the Max–Planck Institute for Radio Astronomy (MPIfR), which coherently sums the digitised signals received from the MeerKAT dishes. Each dish receives a signal with a geometric delay,
\begin{equation}
    \tau_\mathrm{n}(\alpha,\delta,t) = \vec{u}\left( \alpha, \delta,t \right) \cdot \vec{r}_\mathrm{n} / c,
\end{equation}
where $\vec{u}\left( \alpha, \delta \right)$ is the pointing direction, $\vec{r}_\mathrm{n}$ is the location of the antenna, and $c$ is the speed of light.
The channelised complex voltages $\Tilde{f}_n(\nu,t)$ observed by each dish can be coherently combined by accounting for the geometric delays with a complex phase shift, to produce a (channelised total-intensity) TAB, $F(\alpha,\delta,\nu,t)$, via
\begin{equation}
F(\alpha,\delta,\nu,t) = \left|\sum_n \Tilde{f}_n(\nu,t) e^{-2\pi i \nu\tau_n(\alpha,\delta)}\right|^2\,.
\label{e:TAB}
\end{equation}
For an idealised point source at a location $(\alpha_0,\delta_0)$, $\Tilde{f}_n(\nu,t) = A e^{2\pi i \nu \tau_n(\alpha_0,\delta_0,t)}$, where $A$ is a constant amplitude across all dishes. The TAB point-spread function (PSF) is the (fractional) response of Equation~\ref{e:TAB} to this idealised point source as a function of the angular offset from the TAB position,
\begin{equation}
\begin{aligned}
            P\!S\!F(\Delta\vec{u},\nu,t) &= \left|\sum_n e^{2\pi i \nu ( \tau_n(\alpha_0,\delta_0,t) - \tau_n(\alpha,\delta,t))}\right|^2\\
            &= \left|\sum_n e^{2\pi i \nu \Delta\vec{u}(t) \cdot \vec{r}_n} \right|^2\,,
            \end{aligned}
\end{equation}\label{eq:psf}
i.e., it is the 2-dimensional Fourier transform of the antenna locations.
% }

% \textcolor{blue}{
The PSF is both frequency- and time-dependent, as the pointing direction ($\vec{u}$) towards a fixed celestial position changes over time due to Earth's rotation. However, here we are interested in localising rapid transient events, with short time duration, that are detected in band-integrated data. For simplicity, we therefore drop the time- and frequency-dependencies and use instantaneous PSFs at the central observing frequency in the following sections (however, see $\S$\ref{sec:elaborations} for how these effects can be accounted for). 
% }

% \textcolor{red}{
% Beamforming observations with MeerKAT can be performed using the Filterbank and BeamForming User-Supplied Equipment (FBFUSE) compute cluster developed by the Max–Planck Institute for Radio Astronomy (MPIfR), which coherently sums the digitised signals received from the MeerKAT dishes. Each dish receives a signal with a geometric delay,
% \begin{equation}
%     \tau_\mathrm{n} = u\left( \alpha, \delta \right) \cdot r_\mathrm{n}\left( x,y,z \right) / c,
% \end{equation}
% where $u\left( \alpha, \delta \right)$ is the pointing direction, $r_\mathrm{n}\left( x,y,z \right)$ is the location of the antenna, and $c$ is the speed of light. Individual signals can thus be expressed as a function of the delay with respect to some reference time $t_\mathrm{ref}$, i.e.\ $f(t_\mathrm{ref}- \tau_\mathrm{n})$. The point spread function (PSF) of the TAB formed by coherent summation of $N$ signals is then given by
% \begin{equation}
%     P\!S\!F\left( t_\mathrm{ref} \right) = \sum^{\mathrm{N}}_{\mathrm{n}=1} f(t_\mathrm{ref}- \tau_\mathrm{n}) w\left(\tau_\mathrm{n}\right),
% \end{equation}\label{eq:psf}
% where $w\left(\tau_\mathrm{n}\right)$ is the delay correction \citep[][]{1988vanveen}.}

Thus, using the known terrestrial coordinates of the antennas used for a certain MeerKAT observation, the instantaneous PSF can be recovered. The Python package \textsc{Mosaic}\footnote{\url{https://github.com/wchenastro/Mosaic}}\citep{2021chen} derives the MeerKAT TAB PSF given the observation time, target coordinates, antennas included in the array and their relative weighting. The left panel of Fig.~\ref{fig:tilesim} shows the result of a \textsc{Mosaic} PSF simulation using 40 core MeerKAT dishes.

Additionally, \textsc{Mosaic} can use these simulated beam maps to determine the positions of individual TABs such that the desired number of them are optimally hexagonally packed. The right panel of Fig.~\ref{fig:tilesim} shows the circular tiling pattern of 389 TABs formed using the PSF shown in the left panel. The beams are specified to overlap at 25 per cent of their maximum gain\footnote{Wherever "overlap" levels are quoted in this text, unless specified otherwise, they refer to the response level of the TAB PSFs at the centre frequency of the MeerKAT L-band, 1.284~GHz.}, and the semi-axes $a$ and $b$ and position angle $\theta$ marked in the left panel correspond to an ellipse that is a best fit to this level of the PSF.% It is important to note that this tiling simulation rests on two assumptions: first, that the PSF can be approximated by an ellipse at a certain level of response\footnote{While, as mentioned in \citep[][]{2021chen}, this assumption breaks for certain array configurations, it does not actually affect localisation as the PSF is still well modelled.}, and second, that the tiling pattern covers a sufficiently small solid angle that all beams can be approximated by the shape of the TAB centred on the boresight coordinates.

\section{TABLo} \label{sec:tablo}

\begin{figure*}
    \centering
    \includegraphics[width=.9\textwidth]{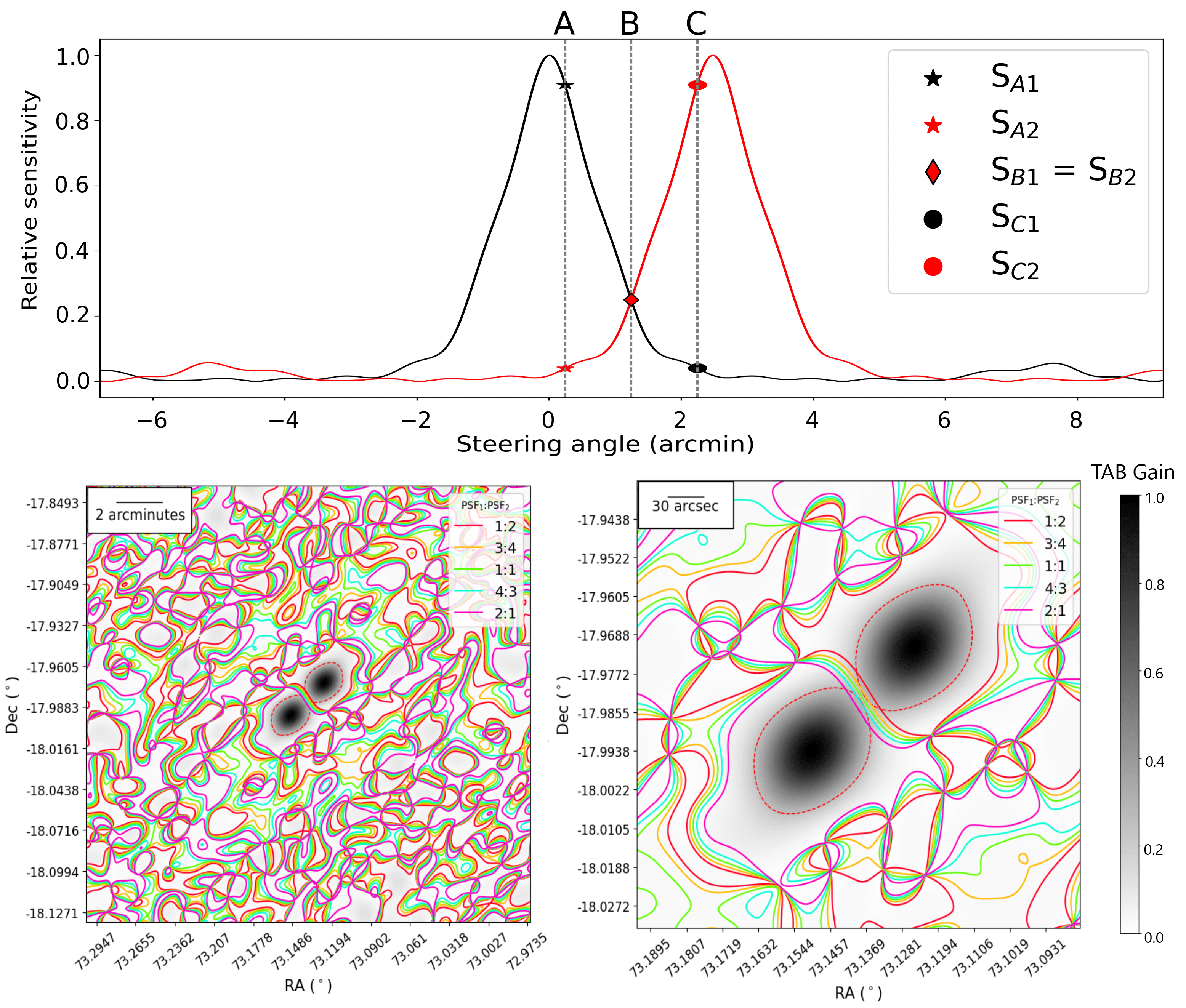}
    \caption{\textbf{Top:} The spatially dependent gain, in normalised units, of two MeerKAT TABs that overlap at the 25 per cent level, modelled using \textsc{Mosaic}. A source at the positions marked A, B, and C would have different observed flux densities in the two beams, as indicated by the red and black markers. \textbf{Bottom:} Two adjacent TABs, with the background grey scale showing the superposition of their respective gains at a given position, and coloured contours shown at different values of their ratio. The right-hand panel is zoomed in on the centres of the beams. Red, dashed ellipses indicate the 25~per~cent levels of the TABs.}
    \label{fig:fluxratio}
\end{figure*}

The method described by \citet{2015obrocka} models the apparent flux density $S_{\nu}$ of a single pulse over space as an intrinsic brightness modulated by the instrumental response (i.e.\ the instantaneous PSF). Since the intrinsic brightness is a constant across TABs, taking the ratio of flux densities in a pair of TABs leaves only the PSFs of those TABs to be modelled using Eq.\ \ref{eq:psf}. The modelled PSF ratios can then be used to predict the observed S/N in adjacent TABs assuming different positions of the source. This idea is illustrated in the top panel of Fig.~\ref{fig:fluxratio}, following \citet[][]{2015obrocka}, in which sources placed at different positions relative to two adjacent TABs' centres have different apparent flux densities. In the bottom panels of Fig.~\ref{fig:fluxratio}, the relative gains of these beams are shown in grey scale, along with coloured contours marking different values of their ratio.

% \begin{figure*}
%     \centering
%     \includegraphics[width=0.82\textwidth]{Figures/2beams.png}
%     \caption{Two adjacent TABs, with the background colour scale showing the superposition of their respective gains at a given position, and contours shown at different values of their ratio. The bottom panel is zoomed in on the centres of the beams. Red, dashed ellipses show the amplitudes at which the beams were specified to overlap, namely 25~per~cent.}
%     \label{fig:beamratios}
% \end{figure*}

Assuming that all TABs are equally sensitive\footnote{In practice, this is not true for TABs with large angular separation because of the effect of the primary beam.}, the measured S/N values will all be proportional to the apparent $S_{\nu}$ of a detected source\footnote{Technically, any observable proportional to the apparent $S_{\nu}$ can be used in this analysis rather than the S/N value, provided the error on the measurement is also known.}. For a source at position $(\alpha,\delta)$, the ratio of S/N values in two TABs in directions $\vec{u}_1$ and $\vec{u}_2$ and with spatially-dependent gain $P\!S\!F_{1}(\alpha,\delta) = P\!S\!F(\vec{u}(\alpha,\delta) - \vec{u}_1)$ and $P\!S\!F_{2}(\alpha,\delta) = P\!S\!F(\vec{u}(\alpha,\delta) - \vec{u}_2)$, respectively, is thus predicted to be
\begin{equation}\label{eq:SN}
    \frac{S\!/\!N_1}{S\!/\!N_2} = \frac{(S_\nu)_1}{(S_\nu)_2} = \frac{P\!S\!F_1(\alpha,\delta)}{P\!S\!F_2(\alpha,\delta)}.
\end{equation}

% Thus, supposing that a source is detected with two S/N values $S\!/\!N_{1}$ and $S\!/\!N_{2}$ in two TABs, and those TABs have modelled PSFs with spatially dependent gain $P\!S\!F_{1}$ and $P\!S\!F_{2}$, then at the position of the source,
% \begin{equation}\label{eq:SN}
%     \frac{P\!S\!F_1}{P\!S\!F_2} = \frac{(S_\nu)_1}{(S_\nu)_2} = \frac{S\!/\!N_1}{S\!/\!N_2}.
% \end{equation}

For each pair of TABs in which a single pulse is detected, \citet[][]{2015obrocka} defines possible positions as those where Eq.~\ref{eq:SN} is true within 1-$\sigma$ errors on the S/N measurements. \citet[][]{2015obrocka} then uses power-law spectral index fitting of the PSFs as an additional constraint on the possible positions. 

TABLo differs from the \citet[][]{2015obrocka} approach in two important ways. Firstly, we fit only the ratios of observed S/N values, and not the differences of spectral indices. Primarily, this is because complex TABs formed over a wide frequency band, such as those of MeerKAT, are generally poorly fit by simple power laws, introducing additional fitting errors that are difficult to account for and often lead to poor localisations. This is discussed further in $\S\ref{sec:spec_ind}$. Secondly, we follow an MLE approach via linear regression analysis. This allows us to derive a positional probability distribution that can be maximised to find the most likely source position, as well as rigorously define standard errors. In contrast to CHIME, which has TABs that span $\sim$0.3 degrees in the North-South direction, MeerKAT's $\sim10$~arcsecond angular resolution is sufficient that the ``gridding'' procedure described in \citet[][]{2021michilli} is not required. Instead, we can make use of S/N values already calculated in formed beams in the process of the transient search; this significantly saves on computing requirements, and allows for real-time localisation. It also enables localisation for surveys that don't have re-phasing capacity. Finally, because TABLo fits the ratio of S/N values in beam pairs, calibrated flux density measurements are not needed, as is the case for the ASKAP method outlined in \citet[][]{2017bannister}.

\subsection{Deriving the probability distribution}
% \textcolor{blue}{
TABLo uses a forward modelling approach: the PSF model produced by Mosaic is used to predict the set of S/N ratios that should be observed between pairs of TABs, as a function of the position of a putative source. These predicted ratios are then compared to the observed S/N ratios. This generates likelihood values for each combination of RA and Dec that can be maximised to find the most likely position, and integrated to calculate uncertainties.
% }

% \textcolor{red}{The MLE method applied to this problem involves fitting the observed S/N ratios to modelled TAB response ratios with RA and Dec as free parameters \citep[see e.g.][]{2018rossi}. This generates likelihood values for each combination of RA and Dec that can be maximised to find the most likely position, and integrated to calculate uncertainties.}

Starting from a list of $Q$ detections of a single pulse above the S/N threshold, we define a vector $S$ containing all independent ratios of their S/N values. Since $S$ is a ratio distribution, it can only be assumed to be Gaussian if the numerator and denominator are themselves independent and normally distributed with a positive mean, and if the error on the denominator is much smaller than its mean \citep{2013diazfrances}. We therefore only use the ratio of the S/N value in each beam to the highest overall S/N (say, S/N$_{1}$). Hence $S$ will be a vector $Q-1$ elements in length,
\begin{equation}\label{eq:S_i}
    S_\mathrm{i} = \frac{S\!/\!N_{\mathrm{i}+1}}{S\!/\!N_{1}}
\end{equation}
with $i = 1,...,Q-1$. Thus, the highest S/N value, $S_{1}$, is always in the denominator in order to prevent deviation from the assumption that the S/N ratios follow a multivariate Gaussian distribution in the limit where $S\!/\!N_{\mathrm{i}+1} \to 0$.

\begin{figure}
    \centering
    \includegraphics[width=0.5\textwidth]{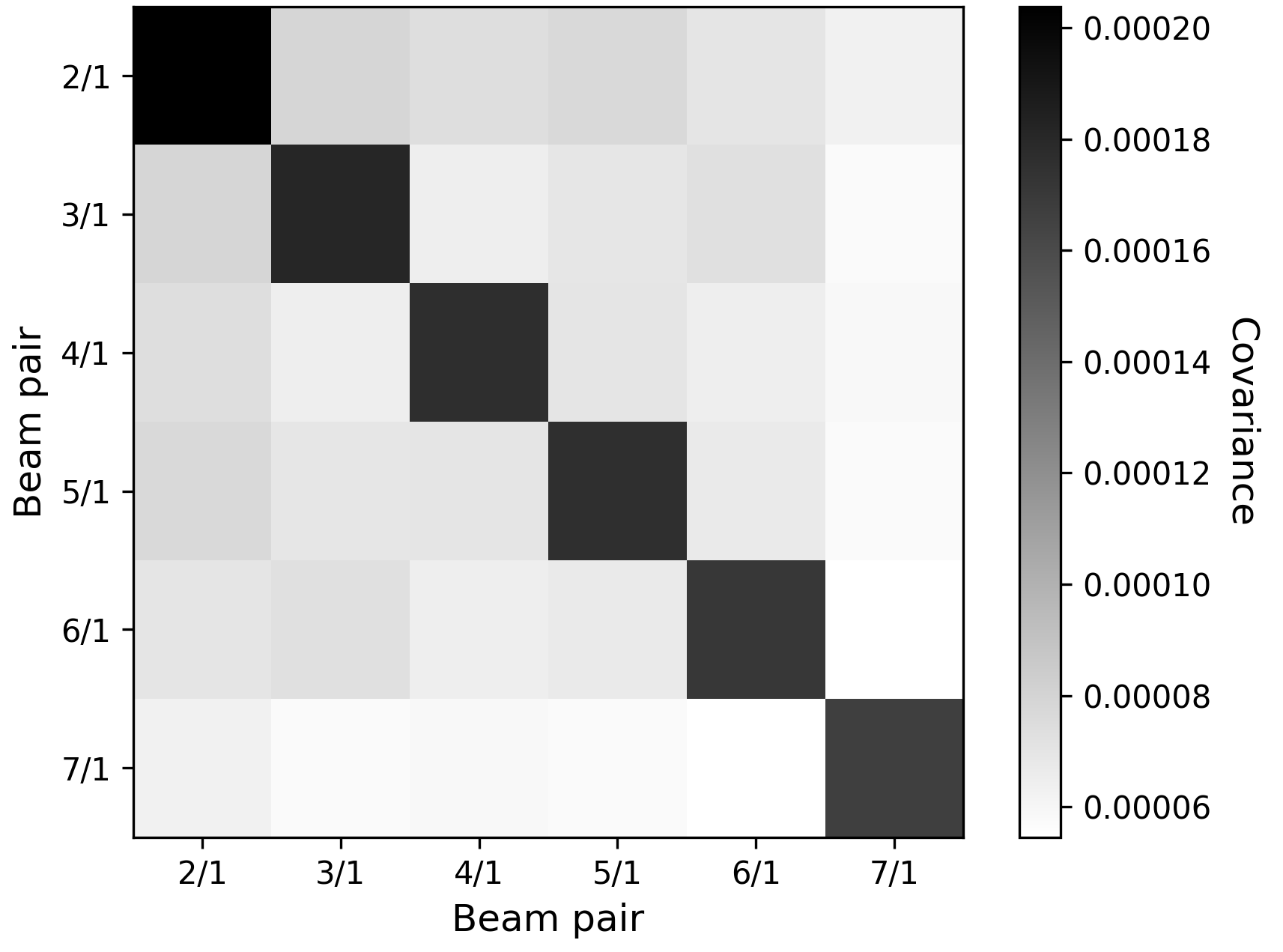}
    \caption{Example covariance matrix of the ratio of the S/N values of a signal measured in seven adjacent TABs at once.}
    \label{fig:covariance}
\end{figure}

S/N values in neighbouring beams will be covariant proportional to the degree to which the TABs are designated to overlap. This effect is small for typical tiling patterns employed in MeerKAT observations. However, the S/N ratios Eq.\ \ref{eq:S_i} are always covariant with one another, as they all share $S/N_1$ as their denominator. In order to compute the covariance of S/N ratios caused by this, we first generate $T$ iterations of randomised S/N values within normally distributed 1-$\sigma$ errors of the observed S/N values\footnote{The covariance matrix could be computed analytically, but the derivation is complicated, and this simulation method is computationally inexpensive.}. For each iteration, we then find $Q-1$ S/N ratios to produce a ($Q-1) \times T$ matrix of S/N ratios. The covariance of $S$ in two beam pairs, $S_\mathrm{i}$ and $S_\mathrm{j}$, is then defined as
\begin{equation}\label{eq:covariance}
    \mathrm{cov}(S_\mathrm{i}, S_\mathrm{j})=\frac{1}{T-1} \sum_{\mathrm{t}=1}^{\mathrm{T}}\left(S_{\mathrm{i},\mathrm{t}}-\bar{S_\mathrm{i}}\right)\left(S_{\mathrm{j},\mathrm{t}}-\bar{S_\mathrm{j}}\right).
\end{equation}
The covariance matrix, $C$, is then a $(Q-1) \times (Q-1)$ matrix defined as
\begin{equation}\label{eq:covariance_matrix}
    C_{\mathrm{i},\mathrm{j}} = \mathrm{cov}(S_{\mathrm{i}},S_{\mathrm{j}}).
\end{equation}
Fig.~\ref{fig:covariance} shows an example covariance matrix for the ratio of S/N values in seven beams, using 1000 iterations of normally distributed random errors. The diagonal elements indicate the variance within distinct beam pairs, while the off-diagonal elements indicate how different pairs of beams covary. 

Next, we model the PSF of each beam using \textsc{Mosaic}, and shift the beam centres to match the listed coordinates as exemplified in Fig.~\ref{fig:fluxratio}. %The PSFs can then be described as $PSF_{\mathrm{i},\mathrm{m},\mathrm{n}}$, where $n= 1,...,N$ and $m=1,...,M$ correspond to a grid of RA and Dec coordinates with a user-specified spacing. 
From this we compute the $Q-1$ predicted ratios between $PSF_{1}$ and every other beam as a function of RA and Dec,
\begin{equation}\label{eq:P_imn}
    \psi_{\mathrm{i}}(\alpha,\delta) = \frac{P\!S\!F_{\mathrm{i}+1}(\alpha,\delta)}{P\!S\!F_1(\alpha,\delta)}.
\end{equation}
We then compute a residual vector $R$ of length $(Q-1)$ as a function of position,
\begin{equation}\label{eq:R_imn}
    R_{\mathrm{i}}(\alpha,\delta) = S_{\mathrm{i}} - \psi_{\mathrm{i}}(\alpha,\delta).
\end{equation}
The generalised least squares method is a technique for performing a linear regression analysis in the case that the model residuals (such as those defined in Eq.~\ref{eq:R_imn}) are correlated to a certain degree \citep[see e.g.][]{1998draper}. This approach defines the probabilistic $\chi^{2}$ metric as the weighted sum of squared deviations, which can be written as a matrix product of the covariance matrix and residuals defined above,
\begin{equation}\label{eq:chi2}
    \chi^{2}(\alpha,\delta) = R^{T} (\alpha,\delta)\, C^{-1}\, R(\alpha,\delta).
\end{equation}
% This can be rewritten in index notation as
% \begin{equation}\label{eq:chi2_matrix}
%     \chi^{2}_{m,n} = \sum_{j} \chi^{2}_{j,m,n} = \sum_{j} \left( R_{j,m,n} \sum_{i} \left(C^{-1}\right)_{i,j} R_{i,m,n} \right).
% \end{equation}

Inserting the residuals from Eq.~\ref{eq:R_imn} and covariance matrix from Eq.~\ref{eq:covariance_matrix} into Eq.~\ref{eq:chi2} allows us to compute the positional likelihood distribution function $\mathcal{L}$ defined by
\begin{equation}\label{eq:loglikelihood}
    \ln\mathcal{L}(\alpha,\delta) = -\dfrac{1}{2}\chi^{2}(\alpha,\delta),
\end{equation}
which can be maximised over the RA and Dec parameters to obtain the most likely coordinates.

In practice, we evaluate the log-likelihood of Eq.~\ref{eq:loglikelihood} over a grid of sky positions. Assuming uniform priors, the probability $Pr$ of the source existing within a given pixel of this grid is approximated simply by normalising $\mathcal{L}(\alpha,\delta)$ so that its sum over all parameter space equals unity, i.e.~
\begin{equation}\label{eq:probability}
    Pr(\alpha,\delta) = A \mathcal{L}(\alpha,\delta),
\end{equation}
where $A$ is a constant such that.
\begin{equation}\label{eq:likelihood_norm}
    A \sum_{\mathrm{m},\mathrm{n}} \mathcal{L}(\alpha_\mathrm{m},\delta_\mathrm{n}) = 1.
\end{equation}

\subsection{Error estimation}

The likelihood can then be used to estimate the statistical uncertainty on the most likely position. If the likelihood follows a two-dimensional Gaussian distribution, it will have the probability density function, in polar coordinates,
% \begin{equation}
% \mathrm{PDF}(r) = \frac{1}{2\pi \sigma^{2}} e^{\frac{-r^{2}}{2\sigma^{2}}},
% \end{equation}
\begin{equation}
\mathrm{PDF}(r) = \frac{1}{2\pi \sigma^{2}} e^{-r^{2}},
\end{equation}
where $\sigma$ is the standard deviation and $r$ is the Mahalanobis distance, i.e.\ the distance from the mean in units of $\sigma$ \citep[see e.g.][]{bensimhoun2009}\footnote{Available online at \url{https://upload.wikimedia.org/wikipedia/commons/a/a2/Cumulative_function_n_dimensional_Gaussians_12.2013.pdf}}. The cumulative distribution function can then be obtained by integrating over $r$ and the azimuth $\theta$, such that
\begin{equation}
\begin{aligned}
        \mathrm{CDF(r)} &= \int^{2\pi}_{0}\int^{r}_0 \frac{r'}{2\pi \sigma^{2}} e^{\frac{-r'^{2}}{2}} dr' d\theta \\ &= 1 - e^{\frac{-r^{2}}{2}}.
\end{aligned}
\end{equation}\label{eq:2d_gaussian}
Points corresponding to integer multiples of the Mahalanobis distance from the mean lie on ellipsoids corresponding to n-$\sigma$ error levels. For example, $\mathrm{CDF}(1) = 39.35$~per~cent of the total probability must be contained within the 1-$\sigma$ uncertainty region. Similarly, the 2-$\sigma$ uncertainty region must contain $\mathrm{CDF}(2) = 86.47$~per~cent of all probability. The two-dimensional probability ${P}_{\mathrm{m},\mathrm{n}}$ in Eq.~\ref{eq:probability} is thus flattened and sorted in descending order of value to form an ordered list $P_{\mathrm{q}}$ with $P_{1} \ge P_{2}...\ge P_{\mathrm{M}\times \mathrm{N}}$. Finding the index $d$ where
\begin{equation}\label{eq:prob_uncertainty}
    \sum_{\mathrm{q}=1}^{\mathrm{d}} P_{\mathrm{q}} = 39.35\%,
\end{equation}
the 1-$\sigma$ uncertainty is specified to include all values $m$ and $n$ for which $P_{\mathrm{m},\mathrm{n}} \le P_{\mathrm{d}}$. Similarly, the 2-$\sigma$ uncertainty is taken so that the sum in Eq.~\ref{eq:prob_uncertainty} equals 86.47~per~cent, while for the 3-$\sigma$ uncertainty it equals 98.89~per~cent.

Note that if the distribution of likelihood deviates from a two-dimensional Gaussian distribution (see $\S\ref{sec:results}$ for examples of such cases), the calculated uncertainties will not strictly correspond to n-$\sigma$ levels, but will instead include equal likelihood as if the distribution were indeed Gaussian.

% The localisation probability over RA and Dec space is referred to as the localisation map, which can be plotted along with contours at the 1-$\sigma$ and 2-$\sigma$ levels of the probability to produce localisation plots such as those in Fig.~\ref{fig:Ideal_test} and others in this paper.

It should also be noted that the statistical uncertainty figures calculated as described in this section are only those arising from the errors on the S/N measurements. Additional errors may stem from imprecise modelling of the PSF. We assume, for instance, that all the antennas have equal gain, and that the phasing is 100~per~cent efficient. Furthermore, the PSF is generated for one instant at one frequency, while the S/N values are for a signal averaged over a wide band (or sub-band; see $\S\ref{subsec:subband}$), and may be averaged over time as well (see $\S\ref{subsec:avg_obs}$). Such errors are not taken into account by the \textsc{SeeKAT} likelihood estimation. For single pulses observed across a relatively narrow band, these errors should be small compared to the error due to the S/N measurement, but other use cases may necessitate a correction coefficient.

\begin{figure*}
    \centering
    \includegraphics[width=\textwidth]{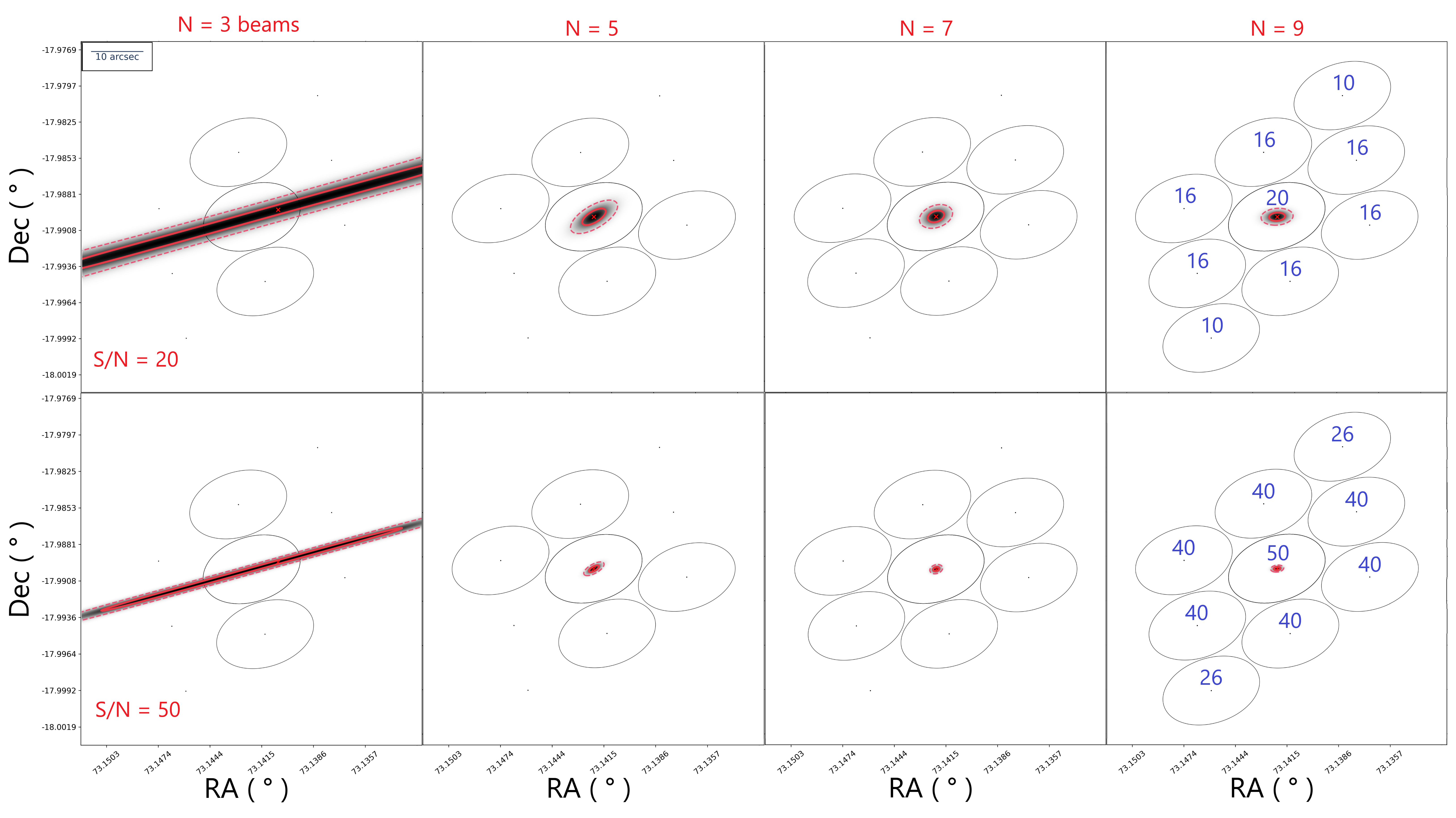}
    \caption{Localisation probability distribution (grey scale) for a simulated ideal case detection in three to nine TABs, assuming a pulse with intrinsic S/N of 20 (top) and 50 (bottom) located at the centre of the middle beam. In each case, the solid red contour represents the 1-$\sigma$ uncertainty level and the dashed red contour represents the 2-$\sigma$ uncertainty level. The S/N values were simulated to correspond to a 20 S/N pulse at the exact centre of the tiling pattern, with beams overlapping at 95~per~cent gain. The blue numbers in the final panels indicate the S/N in each TAB.}
    \label{fig:Ideal_test}
\end{figure*}

\section{Examples and tests} \label{sec:results}
\textsc{SeeKAT}\footnote{\url{https://github.com/BezuidenhoutMC/SeeKAT}} is a Python implementation of the TABLo localisation method developed to localise single pulses found with the MeerTRAP single-pulse detection pipeline \citep[see][]{2018sanidas,2020jankowski,2020malenta,2021rajwade}. In this section, we present ideal-case and real-world tests of the TABLo method using this software.

\subsection{Idealised test cases} \label{subsec:results_ideal}
\begin{figure*}
    \centering
    \includegraphics[width=\textwidth]{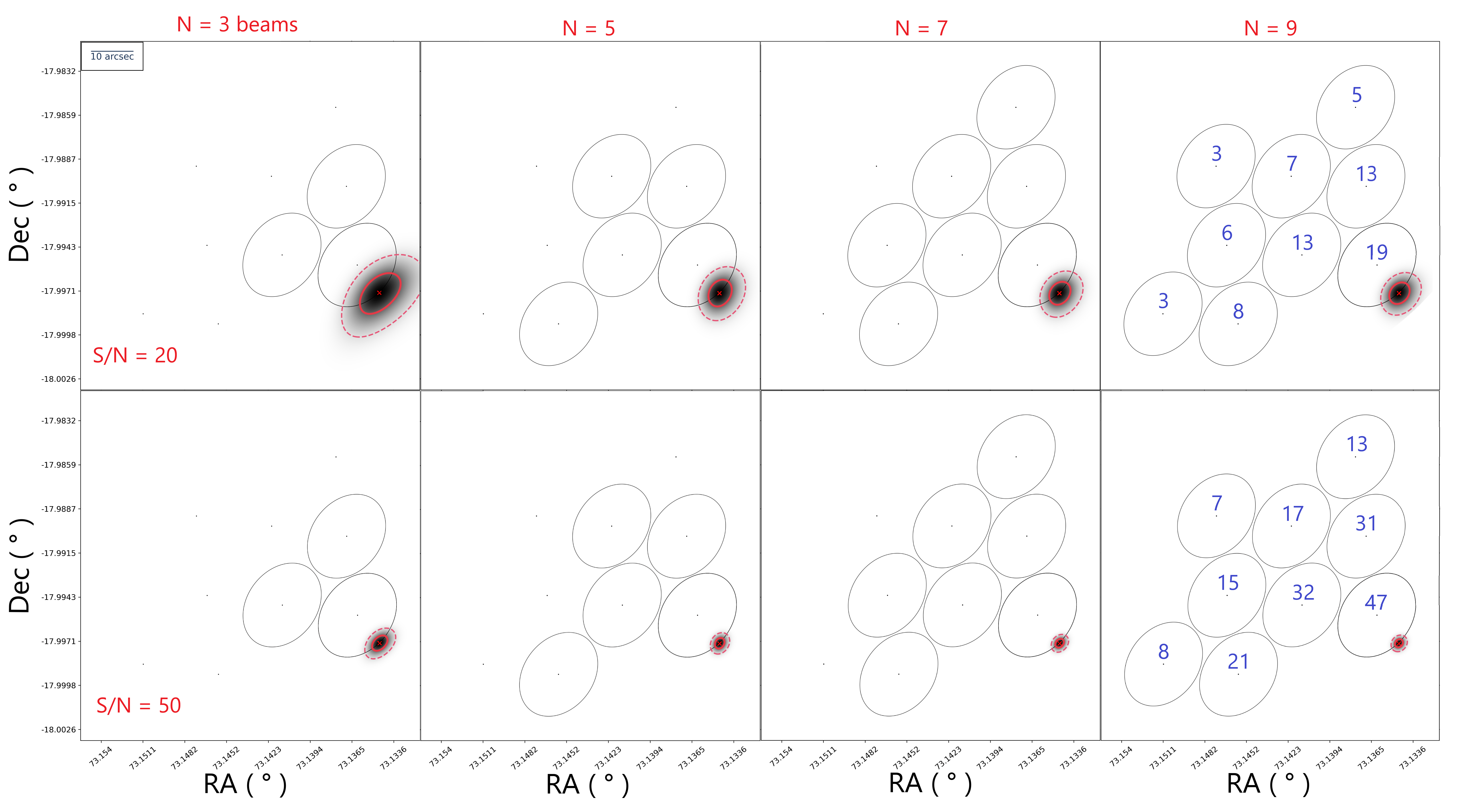}
    \caption{Similar localisation probability distributions as shown in Fig.~\ref{fig:Ideal_test}, but with the source located at the edge of the tiling pattern.}
    %Localisation probability distribution (grey scale) for a simulated ideal case detection in three to ten TABs. In each case the solid red contour represents the 1-$\sigma$ uncertainty level and the dashed red contour represents the 2-$\sigma$ uncertainty level. The S/N values were simulated to correspond to a 20 S/N pulse at the edge of the tiling pattern with beams overlapping at 95~per~cent gain. In each panel the cyan cross indicates most likely position determined by \textsc{SeeKAT}. The red contours indicate the 1-$\sigma$ uncertainty (solid line) and 2-$\sigma$ uncertainty (dashed line)  on the localisation.}
    \label{fig:Ideal_test_offset}
\end{figure*}

\begin{table}
\centering
\begin{tabular}{ccccc}
\hline
 & \multicolumn{2}{c}{S/N = 20} & \multicolumn{2}{c}{S/N = 50}\\ 
\cmidrule(lr){2-3} \cmidrule(lr){4-5} No.\ beams  & Centred pulse & Edge pulse & Centred pulse & Edge pulse\\
\hline
3 & 2100 & 74 & 1000 & 10\\
4 & 32 & 35 & 4.7 & 5.1\\
5 & 17 & 27 & 2.8 & 4.2\\
6 & 9.2 & 26 & 1.5 & 3.8\\
7 & 8.3 & 20 & 1.1 & 3.1\\
8 & 6.9 & 20 & 0.9 & 3.0\\
9 & 5.5 & 19 & 0.8 & 2.7\\

\hline
\end{tabular}
\caption{Areas of the 1-$\sigma$ uncertainty regions, in arcseconds$^{2}$, for test pulses with different positions relative to the tiling centre and different intrinsic S/N values.}\label{tab:ideal_snrs_loc}
\end{table}

To verify that \textsc{SeeKAT} is working correctly given an ideal set of S/N measurements, we tested the code with a simulated bright pulse. First, we generated the TAB PSF for a certain observation along with the beam positions using \textsc{Mosaic}, and then determined the gain of each beam at a particular point. We then multiplied the gain values by the assumed intrinsic S/N to simulate the expected brightness pattern for a pulse at that position. Using the test S/N values derived in this manner as inputs to \textsc{SeeKAT} along with the generated PSF should therefore result in a localisation probability map with a maximum exactly matching the specified position. Any deviation of the most likely coordinates from this position would indicate that the S/N values are not being fit correctly.

% \subsubsection{95~per~cent overlap}

Fig.~\ref{fig:Ideal_test} shows the localisation of such simulated pulses with S/N 20 and 50 that are detected in TABs that overlap at 95~per~cent sensitivity. The pulses are localised using three to nine TABs, successively; gains in localisation precision using more beams than nine were increasingly marginal. We also generated S/N values for a pulse at the edge of the nine beams, 37 arcseconds from the boresight, the localisations resulting from which are shown in Fig.~\ref{fig:Ideal_test_offset}. In both cases the localisation probability distribution is centred exactly on the specified coordinates as expected. The sizes of the 1-$\sigma$ error regions for these pulses using a given number of TABs are listed in Table \ref{tab:ideal_snrs_loc}.

These examples illustrate clearly the three main factors that determine the localisation uncertainty region: the number of TABs in which the pulse is detected, the intrinsic S/N of the pulse, and its position relative to the TAB centres. It is important to note that this case is close to ideal for the purpose of localisation, with an intrinsically bright source detected in multiple very closely spaced beams. For surveys that arrange TABs further apart, the fortuitous placement of the source within the tiling pattern becomes even more significant; a pulse close to the centre of a TAB with no close neighbours may not be detected in enough TABs to be well localised. 

\begin{figure*}
    \centering
    \includegraphics[width=\textwidth]{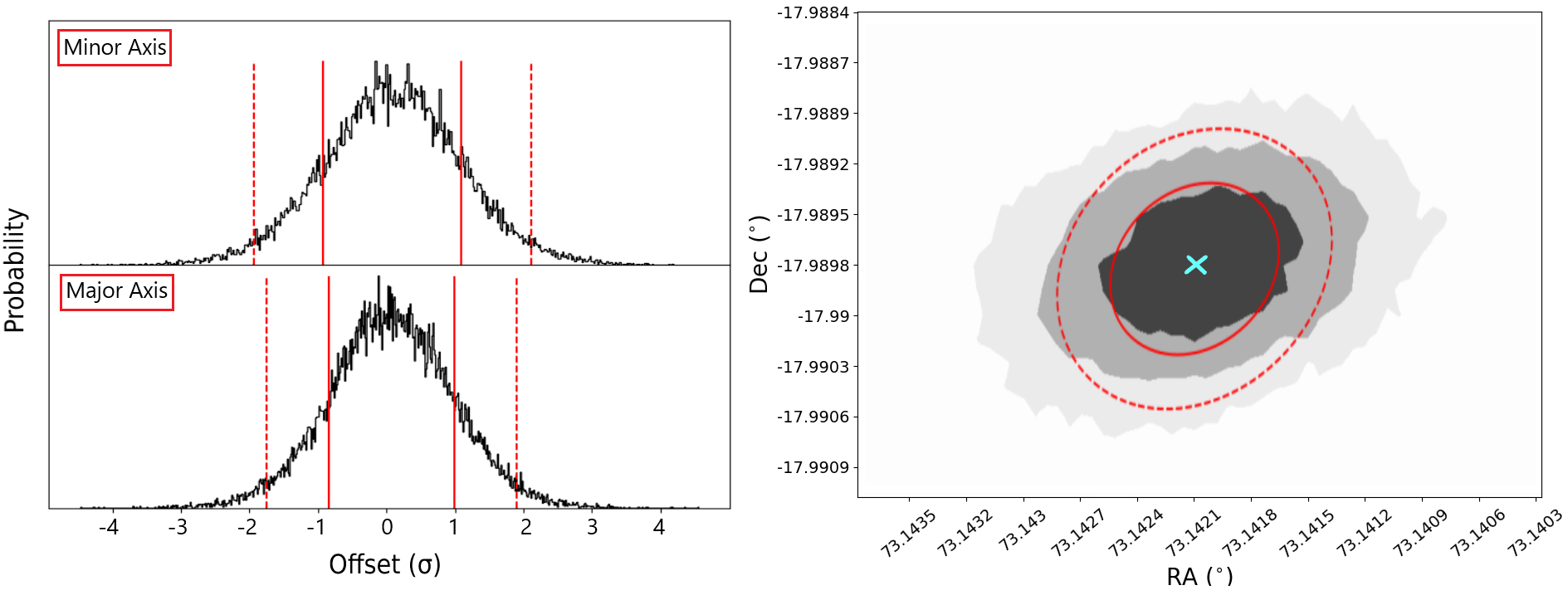}
    \caption{Distribution of localisations for 10000 iterations of a Gaussian random perturbation of the input S/N values. The left-hand panel shows the distribution of offsets along the major (top) and minor (bottom) axis of the localisation region obtained using un-perturbed S/N values (i.e.\ those in Fig.~\ref{fig:Ideal_test}). The right-hand panel contains a heat map of these localisations, with the shading corresponding to 1-, 2-, and 3-$\sigma$ levels. The cyan cross and red ellipses correspond to the most likely position, 1-$\sigma$, and 2-$\sigma$ uncertainty regions of the localisation using un-perturbed S/N values, respectively. The solid and dashed vertical lines indicate one and two standard deviations from the mean, respectively.}
    \label{fig:Ideal_pert}
\end{figure*}

\begin{figure*}
    \centering
    \includegraphics[width=\textwidth]{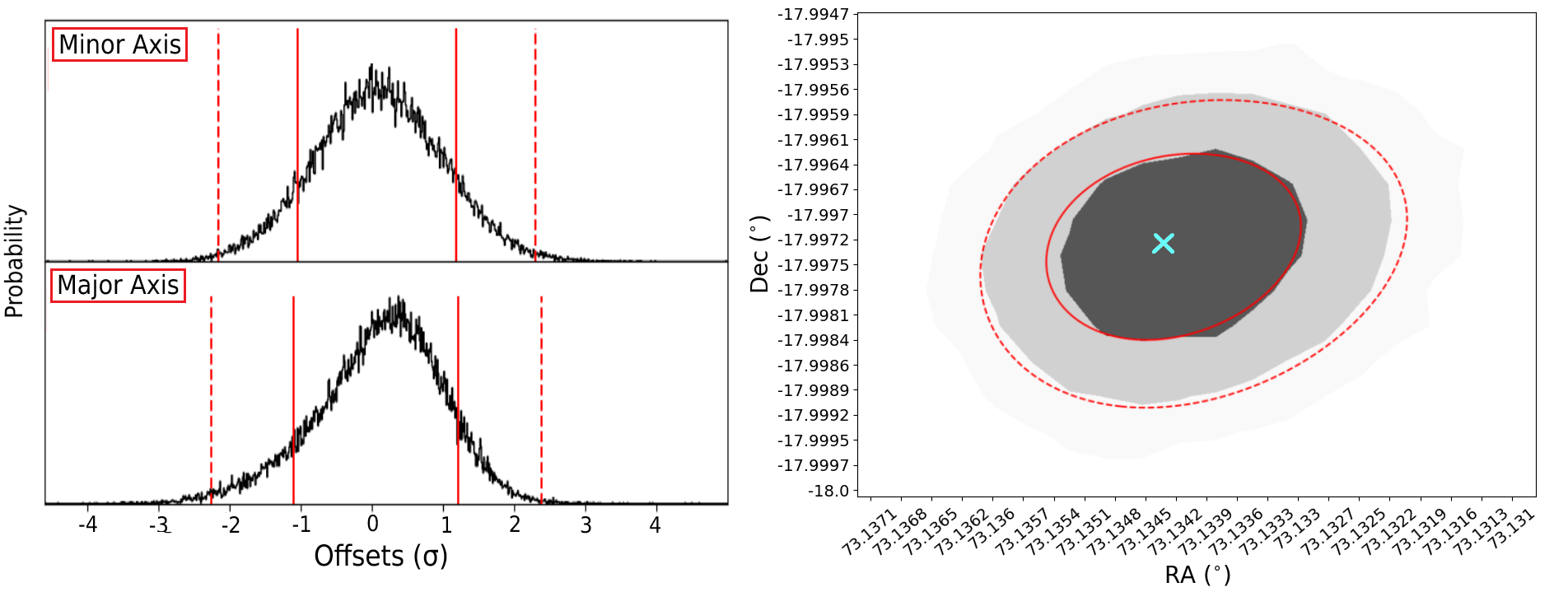}
    \caption{Similar to Fig.~\ref{fig:Ideal_pert}, but for a source located at the edge of the tiling pattern, as shown in Fig.~\ref{fig:Ideal_test_offset}.} 
    
    %Distribution of localisations for 10000 iterations of a Gaussian random perturbation of the input S/N values. The top left panel shows the coordinates of the best-fit position across all 10000 iterations. The top right panel contains a heat map of these localisations, with the greater frequency indicated by the darker shade. In the bottom left panel the distribution of deviations from the localisation using un-perturbed S/N values along the major (top) and minor (bottom) axis of the localisation region is shown. The solid and dashed vertical lines indicate one and two standard deviations, respectively, of the distribution from the mean. The bottom right shows the distributions of perturbed S/N values corresponding to the beams shown in Fig.~\ref{fig:Ideal_test_offset}.}
    \label{fig:Ideal_pert_offset}
\end{figure*}

\subsection{Monte Carlo analysis}

To further test the accuracy of the fitting process, We used the idealised S/N values as in Fig.~\ref{fig:Ideal_test} and Fig.~\ref{fig:Ideal_test_offset} with the addition of Gaussian distributed random perturbations in S/N as inputs for \textsc{SeeKAT}. The noise terms in different TABs were uncorrelated. A number randomly drawn from a Gaussian distribution with a mean of zero and standard deviation of one was added to each of the nine S/N values, and the most likely position calculated by \textsc{SeeKAT} was compared to the position found using the unperturbed values. This process was repeated for 10000 iterations, and the distribution of results compared to the original probability distribution.

The right-hand panel in Fig.~\ref{fig:Ideal_pert} shows a heat map of the 10000 best-fit coordinates determined by \textsc{SeeKAT} using S/N values with a mean of 20 for a pulse at the centre of the tiling pattern as in Fig.~\ref{fig:Ideal_test}. The best-fit positions are normally distributed around that of the un-perturbed case, with standard deviations closely matching those of the original fit, shown in red. 

In the left-hand panel of Fig.~\ref{fig:Ideal_pert}, offsets of the perturbed fits along the major (top) and minor (bottom) axes of the original fit are plotted. The offsets along both axes are normally distributed around a mean of zero. These distributions also have one and two standard deviations (indicated by solid and dashed vertical lines, respectively), that align closely with the 1- and 2-$\sigma$ uncertainties for the original fit indicated on the horizontal axes. The probability distribution generated by \textsc{SeeKAT} therefore agrees well with that predicted by a Monte Carlo iterative approach. Fig.~\ref{fig:Ideal_pert_offset} shows the result of the same analysis for the case where the source is at the edge of the tiling pattern.

\begin{table}
\centering
\begin{tabular}{c|ccc}
\hline
& Mean & 1-$\sigma$ error & 2-$\sigma$ error\\
\hline
% Centred pulse & 0.07 & 0.09 & 0.93 & 1.02 & 1.86 & 2.06\\
Centred pulse & 0.11 & 0.07 & 0.12\\
Edge pulse & 0.08 & 0.17 & 0.38\\
\hline
\end{tabular}
\caption{Offsets, in units of $\sigma$, of the probability distribution determined using \textsc{SeeKAT} compared to that found by Monte Carlo analysis, for two 20 S/N pulses at the centre and edge of the tiling pattern, respectively.}\label{tab:montecarlo_95}
\end{table}

The offsets between the probability distributions predicted using \textsc{SeeKAT} and the Monte Carlo analysis are compiled in Table~\ref{tab:montecarlo_95}. The quoted offsets are defined as the root sum squared (RSS) of differences in the minor and major axis directions, in units of $\sigma$. While in all cases the offsets are smaller than unity, \textsc{SeeKAT}'s 1-$\sigma$ and 2-$\sigma$ estimates for the edge pulse are mildly underestimated. An important factor is that the localisation probability shown in Fig.~\ref{fig:Ideal_test_offset} is somewhat asymmetrical, particularly in the direction of the major axis of the unperturbed uncertainty region, and therefore not exactly normally distributed. An ellipse is therefore unlikely to be well-fit to the 1-$\sigma$ error on the distribution, leading to errors on the measured offsets.

\subsection{Tests with sparse spatial sampling }

% \begin{table*}
% \centering
% \begin{tabular}{cccccccccc}
% \hline
%  & \multicolumn{3}{c}{TAB 1} & \multicolumn{3}{c}{TAB 2} & \multicolumn{3}{c}{TAB 3}\\
% \cmidrule(lr){2-4}\cmidrule(lr){5-7}\cmidrule(lr){8-10}
% Overlap level & RA (hms) & Dec (dms) & S/N & RA (hms) & Dec (dms) & S/N & RA (hms) & Dec (dms) & S/N\\
% \hline
% 25~per~cent & 04:52:27.8 & $-$17:59:36.2 & 8.73 & 04:52:28.5 & $-$17:58:05.6 & 8.20 & 04:52:34.1 & $-$17:59:23.4 & 8.71\\
% 50~per~cent & 04:52:30.3 & $-$17:58:27.9 & 20.66 & 04:52:29.6 & $-$17:59:31.8 & 18.48 & 04:52:34.1 & $-$17:59:23.4 & 20.61\\
% \hline
% \end{tabular}
% \caption{Coordinates and idealised S/N values for three TABs in which a single pulse with an intrinsic S/N of 50 was detected.}\label{tab:ideal_25_50_snrs}
% \end{table*}

The previously described scenarios correspond to a situation where the approximate coordinates of an already discovered source is re-observed with very tightly packed TABs specifically to maximise the potential for a precise \textsc{SeeKAT} localisation. However, in the normal operation of an untargeted transient and pulsar search, surveys are unlikely to tile their TABs so closely, and instead will tend to maximise spatial coverage of the TABs while ensuring sufficient sensitivity to detect a reasonably bright pulse anywhere within the tiling. For the MeerTRAP real-time transient and pulsar search, for instance, TABs are normally arranged to intersect at to overlap at 25~per~cent, while for the (mostly) targeted TRAPUM MeerKAT transient and pulsar search project \citep[][]{2016stappers}, they are variably tiled with overlaps in the 30~per~cent to 80~per~cent range depending on the target. 

\begin{figure*}
    \centering
    \includegraphics[width=\textwidth]{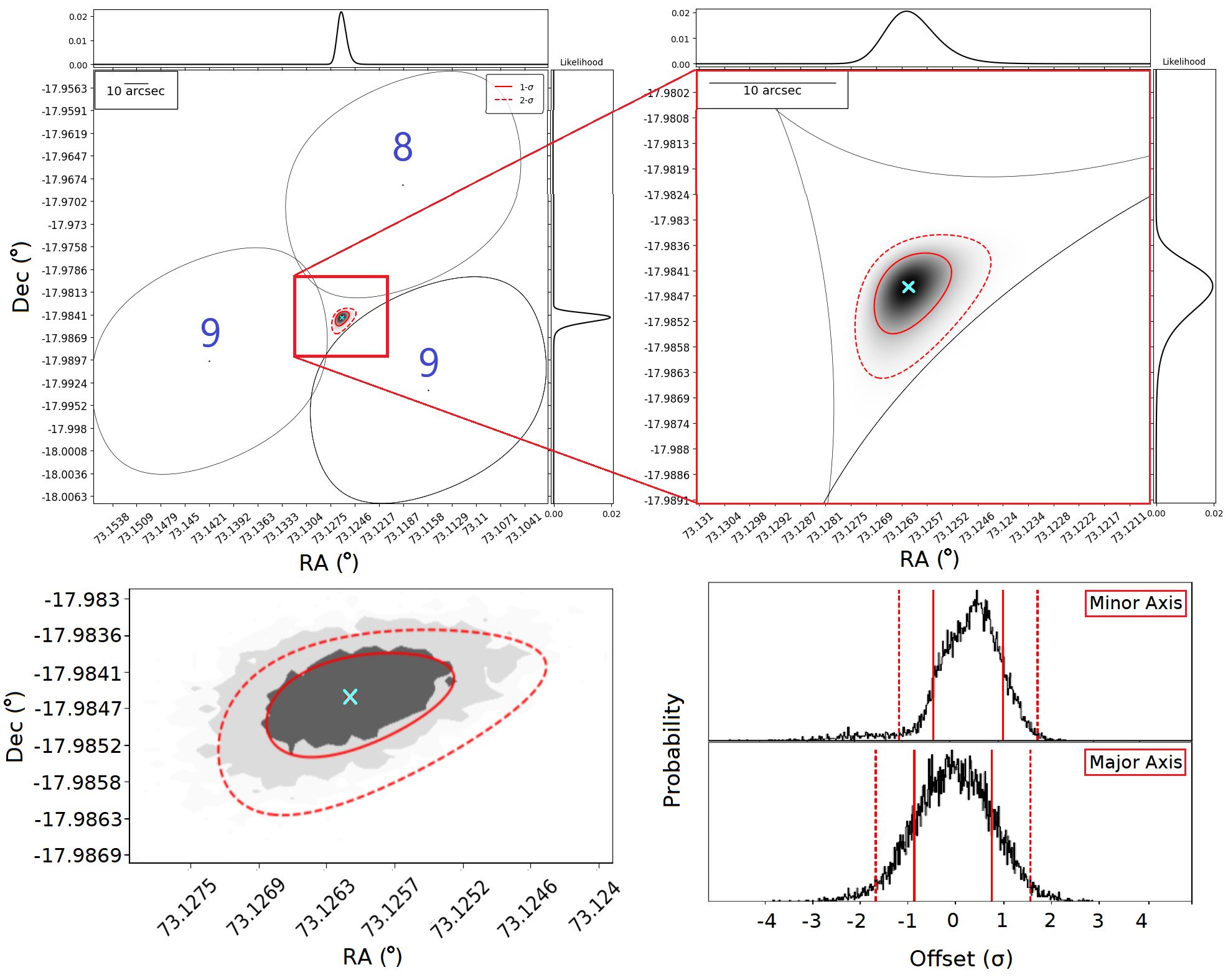}
    \caption{Localisation of a single pulse located at RA 04:52:34 and Dec $-$17:59:23 and detected in three adjacent beams that intersect at 25~per~cent gain. The top panels show the three TABs and the likelihood distribution determined using \textsc{SeeKAT}. The bottom-left panel shows a heatmap of localisations resulting from a random Gaussian perturbation analysis, with the cyan cross and red ellipses indicating the best-fit position, 1-$\sigma$, and 2-$\sigma$ uncertainty regions determined using the un-perturbed S/N values, respectively. In the bottom-right panels, the distributions of offsets of the localisations using perturbed S/N values in the minor axis and major axis directions, respectively, are plotted.}
    \label{fig:ideal_25pc}
\end{figure*}

\begin{figure*}
    \centering
    \includegraphics[width=\textwidth]{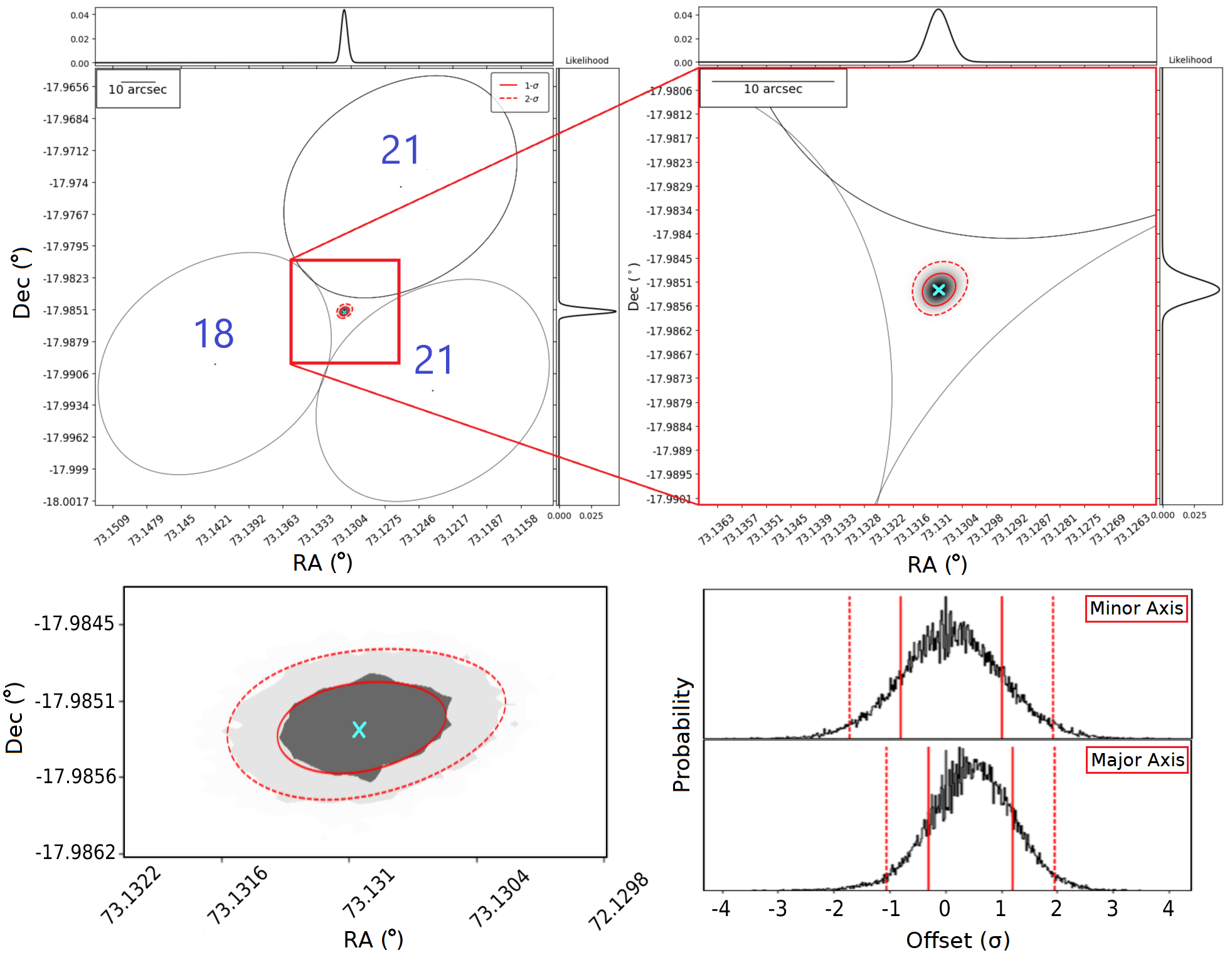}
    \caption{Similar to Fig.~\ref{fig:ideal_25pc}, but for beams that intersect at 50~per~cent of their maximum gain.} 
    % Localisation of a single pulse in three adjacent beams that intersect at 50~per~cent of their maximum gain. The top panels show the three TABs and the likelihood distribution determined using \textsc{SeeKAT}. The middle-left and bottom-left panels show the distribution of all localisations and heatmap of localisations resulting from a random Gaussian perturbation analysis, respectively. In the middle-right and bottom-right panels the distributions of offsets of the localisations using perturbed S/N values in the minor axis and major axis directions, respectively, are plotted.}
    \label{fig:ideal_50pc}
\end{figure*}

Using the same observational parameters as in the previous section, we created idealised S/N values for TABs tiled at 25 and 50~per cent overlap. For a TAB configuration with 25~per~cent overlap, a pulse situated exactly in the middle of three beams must have a minimum S/N value of about 50 for it to be detectable in all three above a cut-off S/N of eight. To perform a Monte Carlo analysis on these more sparsely-sampled configurations, we therefore generate S/N values with a mean of 50.

The resulting \textsc{SeeKAT} localisations are shown in the top panels of Fig.~\ref{fig:ideal_25pc} (25~per~cent) and Fig.~\ref{fig:ideal_50pc} (50~per~cent), respectively. The bottom panels in each figure show the results of a Monte Carlo analysis, as in the previous section. In Table~\ref{tab:montecarlo_25_50}, we list the sizes of the 1-$\sigma$ localisation region for each localisation. We also collate the offsets between the probability distributions determined using TABLo compared to the Monte Carlo analysis.

\begin{table}
\centering
\begin{tabular}{ccccc}
\hline
 & & \multicolumn{3}{c}{Offsets, TABLo vs Monte Carlo}\\
  \cmidrule(lr){3-5}
Overlap & 1-$\sigma$ area & Mean & 1-$\sigma$ error & 2-$\sigma$ error\\
\hline
% Centred pulse & 0.07 & 0.09 & 0.93 & 1.02 & 1.86 & 2.06\\
25~per~cent & 28.29 & 0.27 & 0.34 & 0.71\\
50~per~cent & 5.73 & 0.44 & 0.26 & 0.52 \\
\hline
\end{tabular}
\caption{Area (in arcsecond$^{2}$) of the 1-$\sigma$ uncertainty region of localisations of a 50 S/N pulse in three TABs, as well as offsets (in units of $\sigma$) between the probability distribution determined using TABLo compared to that determined from a Monte Carlo analysis.}\label{tab:montecarlo_25_50}
\end{table}

The probability distributions for these localisations are noticeably irregularly shaped. For the 25~per~cent example, the distribution of offsets has a strong tail in the negative minor axis direction, causing the TABLo and Monte Carlo 2-$\sigma$ uncertainty estimates to differ by as much as 0.71 $\sigma$. Note, however, that this comparison does not take into account the errors on the Monte Carlo estimates themselves, exaggerating the discrepancy. For the 50~per~cent overlap case, meanwhile, the TABLo localisation is close to normal in the minor axis direction, and therefore matches the Monte Carlo distribution well; the major axis direction, however, is significantly skewed in the positive direction, causing their mean values to vary by 0.44 $\sigma$. 

In contrast with the examples in the previous section, which illustrate how well the \textsc{SeeKAT}-determined likelihood matches the expected distribution in ideal or near-ideal circumstances with detections in many TABs, the examples in the 50~per~cent and, especially, the 25~per~cent overlap cases, show the performance of \textsc{SeeKAT} in the opposite edge case. i.e.~the faintest single pulse detectable in three distant TABs. In both instances, the means of the localisations using perturbed S/N values are slightly offset from the most likely unperturbed position in one direction (although both are within 0.5-$\sigma$). Additionally, the offsets in both cases have long tails that represent a departure from the normal distribution, which likely has a sizable impact on determining the 1-$\sigma$ and 2-$\sigma$ uncertainty regions.
\begin{figure*}
    \centering
    \includegraphics[width=\textwidth]{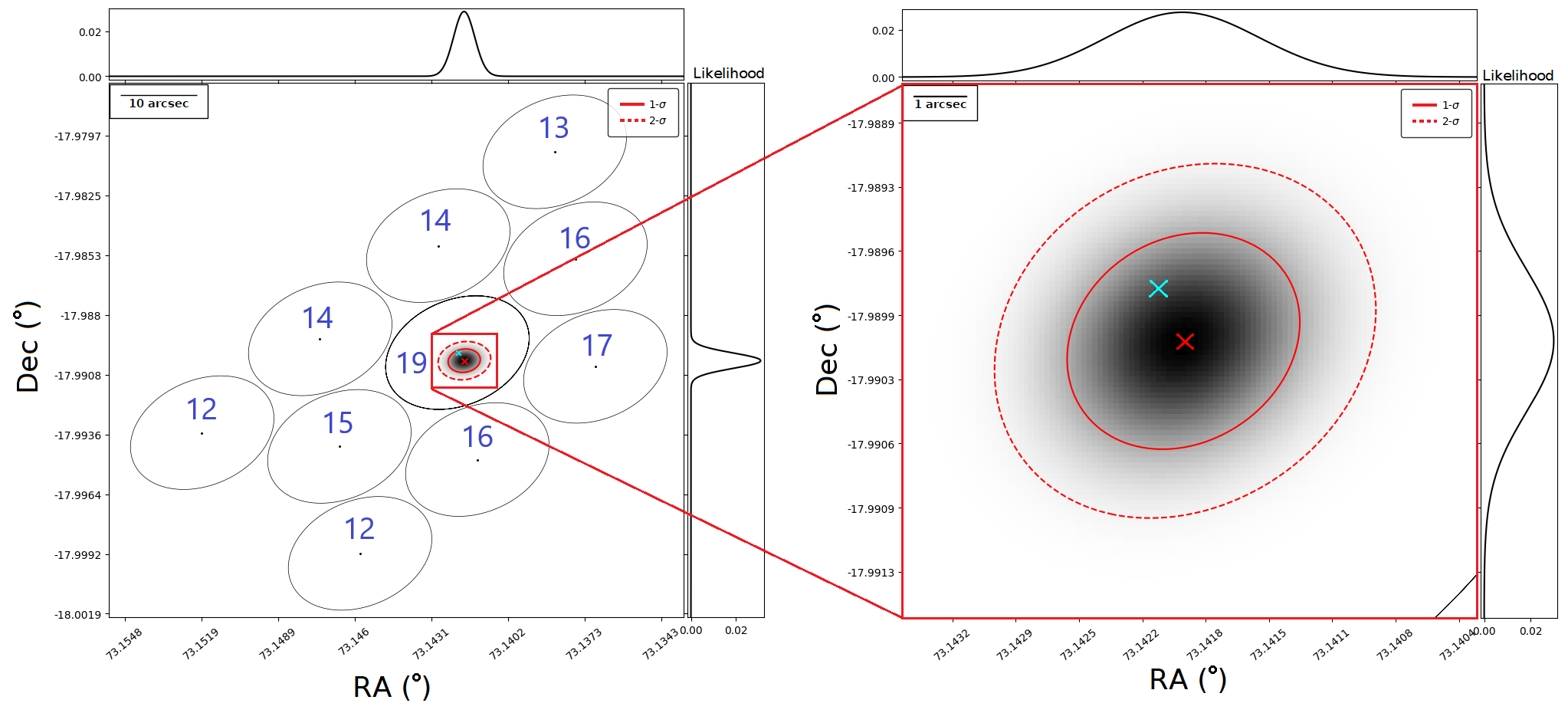}
    \caption{\textsc{SeeKAT} localisation of PSR~B0450$-$18 using S/N values measured during a MeerKAT observation centred on the source. The right panel is zoomed in on the localisation region. The red cross shows the most likely position determined by \texttt\textsc{SeeKAT}, while the cyan cross shows the pulsar's known coordinates, the uncertainties on which are too small to be seen.}
    \label{fig:B0450_loc}
\end{figure*}

Based on these results, we recommend using localisations from \textsc{SeeKAT} judiciously in the case of a few detections in disparate TABs. In such circumstances, users may be better served to quote the 2-$\sigma$ reported by \textsc{SeeKAT} rather than the 1-$\sigma$ uncertainty.

\subsection{Real-world examples}\label{subsec:real_examples}
\subsubsection{PSR~B0450$-$15}
A MeerKAT observation on 16 July 2020 centred on the bright pulsar PSR~B0450$-$15 produced detections above the S/N threshold in 10 adjacent TABs, which overlapped at 95~per~cent gain. The brightest detection, with a S/N\footnote{Unless otherwise specified, all measured S/N values were determined using the \textsc{spyden} Python package, available at \url{https://bitbucket.org/vmorello/spyden/}.} of 19, was in the central TAB of the tiling pattern. The \textsc{SeeKAT} localisation, with best-fit coordinates of RA 04:52:34.06 and Dec $-$17:59:24, is shown in Fig.~\ref{fig:B0450_loc}. The 1-$\sigma$ and 2-$\sigma$ uncertainty regions are roughly circular, with half-widths of about 2~arcseconds and 3~arcseconds, respectively. The known coordinates for PSR~B0450$-$18 are RA 04:52:34.1057(1) and Dec $-$17:59:23.371(2), which is an angular distance of 1.08 arcseconds from the most likely coordinates determined using \textsc{SeeKAT}.  The known coordinates are therefore within the 1-$\sigma$ error region predicted by \textsc{SeeKAT}.
\subsubsection{PSR~J1843$-$0757}

\begin{figure*}
    \centering
    \includegraphics[width=\textwidth]{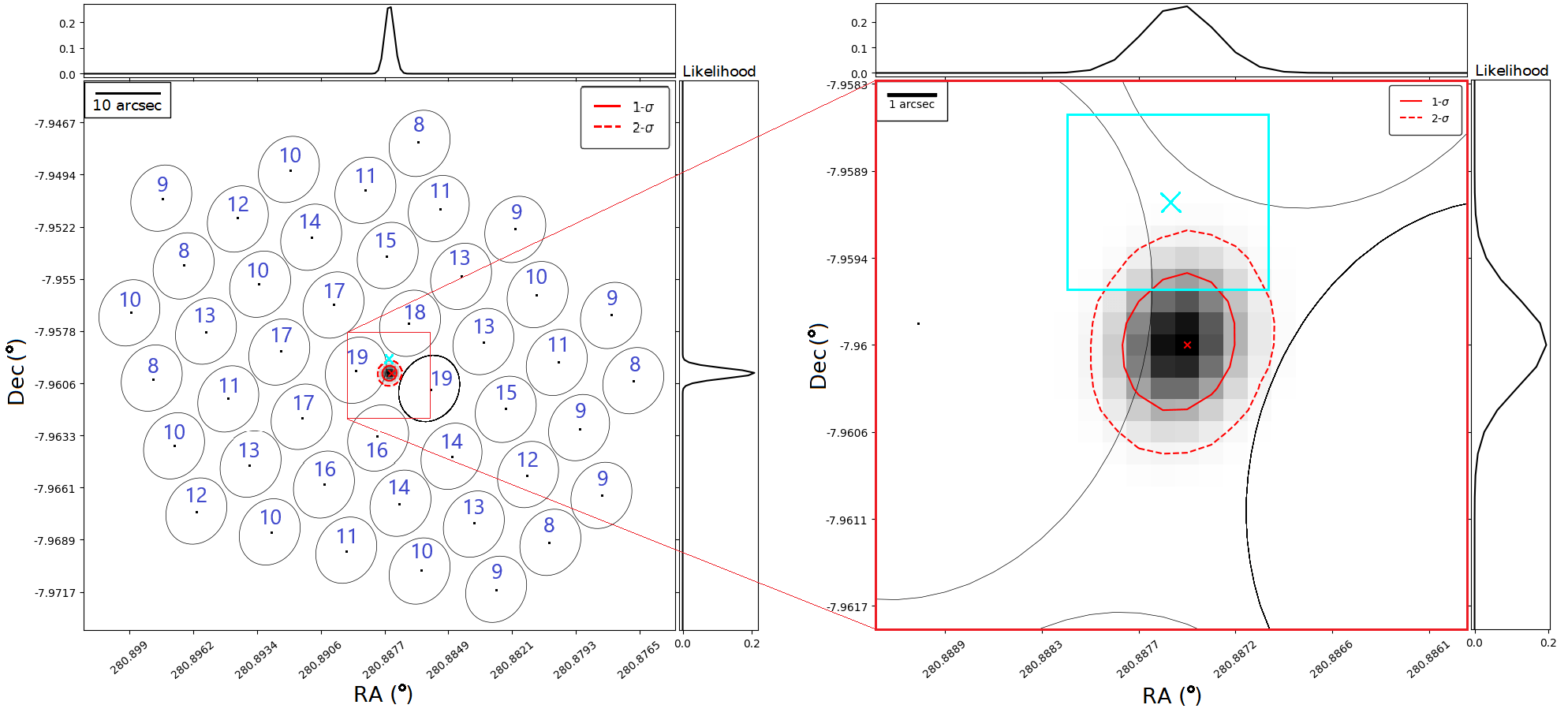}
    \caption{\textsc{SeeKAT} localisation of PSR~J1843$-$0757 using S/N values measured in 44 MeerKAT TABs. The red cross indicates the most likely position determined by \textsc{SeeKAT}, with the red contours the 1-$\sigma$ and 2-$\sigma$ uncertainty on this position, and the cyan cross shows the localisation derived through timing of the source. The right panel is zoomed in on the localisation region. The cyan box approximates the $1$-sigma errors on the timing localisation. Note that the sides of the box correspond to the one-dimensional marginalisation of the true timing uncertainty, which is elliptical. The \textsc{SeeKAT} and timing localisations coincide within the 1-$\sigma$ uncertainty level.}
    \label{fig:MTP1_loc}
\end{figure*}

Another example of \textsc{SeeKAT} in use is presented in \citet[][]{2022bezuidenhout}. A new pulsar, PSR~J1843$-$0757, was discovered by MeerTRAP's real-time single-pulse detection pipeline. In a follow-up observation of the source with MeerKAT, the TABs were tiled to overlap at their 98 per cent level. Using a single pulse detected in 44 TABs at once, the source was localised to RA 18:43:33.01 and Dec $-$07:57:36 with 1-$\sigma$ statistical uncertainty of about 1 arcsecond. A coherent timing solution for the source gives a best-fit position of RA 18:43:33.06(2) and Dec $-$07:57:33(2), which is 3 arcseconds from the \textsc{SeeKAT} position. Fig.~\ref{fig:MTP1_loc} shows the \textsc{SeeKAT} localisation of this source. In the right-hand panel, which is zoomed in on the localisation region, one can see that the \textsc{SeeKAT} and timing localisations are consistent to within the 1-$\sigma$ uncertainty level.

\subsubsection{Globular cluster pulsars}

% \begin{figure}
%     \centering
%     \includegraphics[width=0.5\textwidth]{Figures/NGC6440G_localisation_LVC.png}
%     \caption{Localisation probability for the globular cluster pulsar NGC~6440G. The red cross indicates the most likely position determined by \textsc{SeeKAT}, and the red contours show the 1-$\sigma$, 2-$\sigma$, and 3-$\sigma$ uncertainty on this position. The cyan cross shows the localisation derived through timing of the source. The \textsc{SeeKAT} and timing localisations coincide within the 3-$\sigma$ uncertainty level.}
%     \label{fig:NGC6440G_loc}
% \end{figure}

A collaboration of the MeerTime \citep{2020bailes} and TRAPUM MeerKAT LSPs targeted the globular cluster (GC) NGC~6440 with MeerKAT to search for and time pulsars \citep[][]{2022vleeschower}. By tiling 288 TABs to overlap at 70~per~cent gain, two new pulsars were discovered. For one pulsar, NGC~6440G, however, the MeerKAT timing solution could not initially be phase-connected with that obtained using archival Green Bank Telescope (GBT) observations. Because of the weak nature of this pulsar, an accurate localisation of the source was needed for sufficiently accurate timing to phase-connect the ToAs. 

Hence, \textsc{SeeKAT} was used with MeerKAT detections of NGC~6440G over two epochs in four and seven TABs, respectively, to localise the source to the coordinates RA 17:48:52.76 and Dec $-$20:21:38.45. This localisation allowed for the GBT and MeerKAT ToAs to be phase-connected, and the resulting long-term timing solution gave a position of RA 17:48:52.6460(4) and Dec $-$20:21:40.63(1), 2.5 arcseconds from the most likely \textsc{SeeKAT} position. The timing solution in this case is consistent with the \textsc{SeeKAT} localisation within the 3-$\sigma$ uncertainty level. This relatively large offset is likely a reflection of the fact that the S/N values were averaged over long observations of up to four hours, during which time the TAB orientation can change by up to 60$^{\circ}$. See $\S\ref{subsec:avg_obs}$ for a discussion of this problem.

\textsc{SeeKAT} has also been used to localise pulsars discovered in other MeerKAT GC surveys, e.g.\ NGC~6624 \citep[][]{2022abbate} and NGC~1851 \citep[][]{2022ridolfi}. At the time of writing, however, no independent localisations of these sources are available for comparison.

\subsubsection{FRB~20210123}

FRB~20210123 was discovered by MeerTRAP on 23 January 2021. A single pulse from the source was detected in the MeerKAT incoherent beam (IB) with S/N 11.8, as well as in two TABs, with S/N 22.8 and S/N 8.83. The TABs were tiled to overlap at their 25 per cent level. Details of this discovery will be presented in a forthcoming paper. In order to localise the FRB with \textsc{SeeKAT}, the disparity in the gain of the IB and TABs needed to be taken into account.

The ratio of the gain of the centre of a TAB G$_{\mathrm{TAB}}$ to the gain of the IB at that position G$_{\mathrm{IB}}$ is given by
\begin{equation}\label{eq:IB_ratio}
    \frac{G_{\mathrm{TAB}}}{G_{\mathrm{IB}}} = \frac{N_{\mathrm{TAB}}}{\sqrt{N_{\mathrm{IB}}}},
\end{equation}
where N$_{\mathrm{TAB}}$ and N$_{\mathrm{IB}}$ are the number of antennas that were used to synthesise the TABs and IB, respectively. For the observation during which FRB~20210123 was discovered, N$_{\mathrm{TAB}} = 40$ and N$_{\mathrm{IB}} = 60$. \textsc{SeeKAT} was modified such that Eq.~\ref{eq:P_imn} was replaced by
\begin{equation}
    \psi_{\mathrm{i},\mathrm{m},\mathrm{n}} = \frac{G_{\mathrm{IB}} N_{\mathrm{TAB}}}{\sqrt{N_{\mathrm{IB}}}} \frac{PSF_{\mathrm{i}+1,\mathrm{m},\mathrm{n}}}{PSF_{\mathrm{IB},\mathrm{m},\mathrm{n}}},
\end{equation}
where G$_{\mathrm{IB}}$ is in each case the gain of the IB PSF at the centre of the TAB.

\begin{figure}
    \centering
    \includegraphics[width=.49\textwidth]{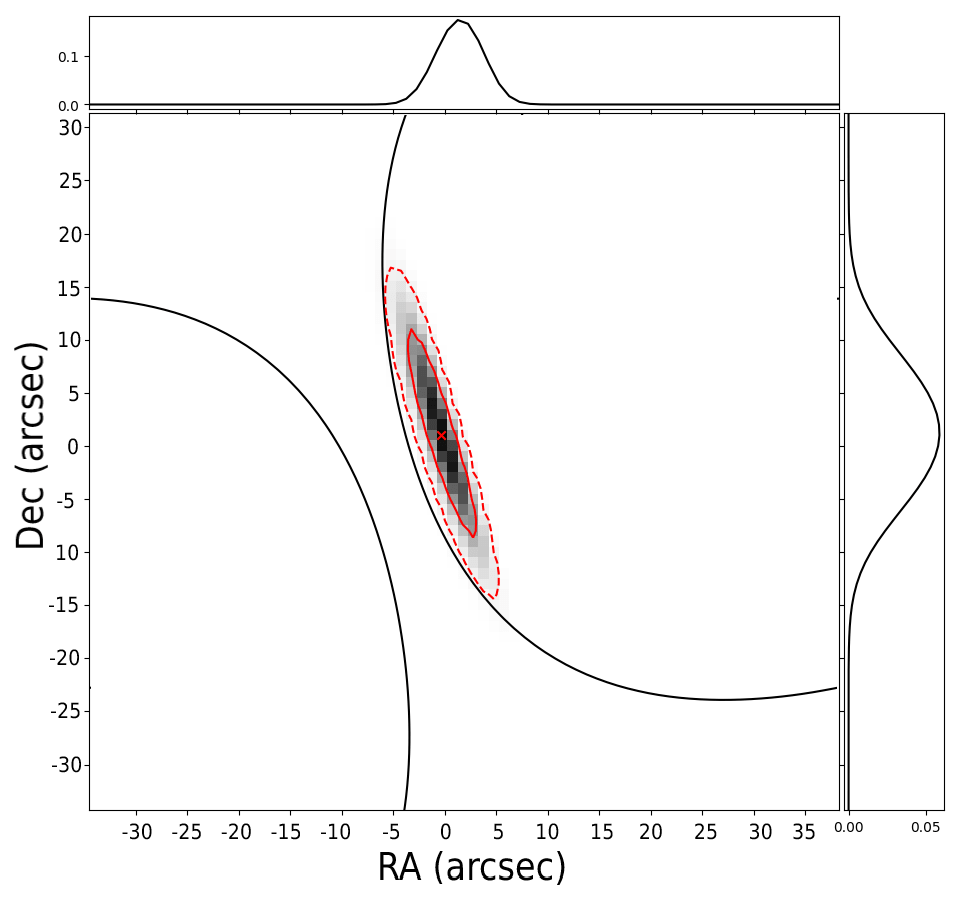}
    \caption{Localisation probability FRB~20210123, which was localised with \textsc{SeeKAT} via a single pulse detected in the MeerKAT IB and two adjacent TABs.}
    \label{fig:FRB8_loc}
\end{figure}

The resulting localisation is shown in the top panel of Fig.~\ref{fig:FRB8_loc}. The most likely position was calculated along with an elliptical 1-$\sigma$ uncertainty region with a half width of $1.5\times9$~arcseconds.

% -----------------------------------------------------------SECTION 5---------------------------------------------------------------------------------------------------------------

\section{Additional methods}\label{sec:elaborations}

The MLE approach to TABLo put forward in $\S\ref{sec:tablo}$ opens up the possibility of incorporating additional information about the pulse detection to further improve the localisation. In this section, we will list various ways that the approach to TABLo presented in this work may be improved upon, as well as the challenges or impediments of those elaborations on the method. Bezuidenhout (in prep.)\footnote{PhD thesis soon online; available upon request.} contains a more detailed evaluation of these approaches.

\subsection{Sub-band localisation}\label{subsec:subband}

A drawback of the approach described in the previous section is that the measured S/N is for a pulse integrated over a wide band, and that a single PSF is generated at a certain frequency despite the beam shape being highly frequency dependent. It may thus be preferable to perform fits in various frequency sub-bands, and add the resulting log-likelihood distributions together to produce a combined localisation. The pulse spectrum may also be used to weight the localisation to reduce the effect of frequencies where the S/N is low. Since side-lobes are significantly frequency-dependent, this method may be better equipped to discriminate side-lobe detections. On the other hand, the signal in individual sub-bands may be diluted to the overall detriment of the fit. Preliminary testing has indicated that this approach may produce more precise localisations in some cases, but its accuracy depends strongly on the strategy used to divide the frequency band.

\subsection{Time-averaged localisation of repeating sources}\label{subsec:avg_obs}
While designed to localise single pulses, TABLo can be applied without much modification to time-averaged observations of repeating sources like pulsars and some fast transients. In this case, average S/N values in each TAB can be used in concert with representative PSFs. However, there is the added complication of accounting for the sometimes drastic evolution of the PSF over the course of the observation \citep[see the discussion around this in][]{2021chen}. Time-averaged PSFs may be used, but they would become less representative for longer observations. A potentially superior approach is viable for sources that are visible over multiple sub-integrations, in which case log-likelihood maps for individual sub-integrations can be combined, similar to the sub-banding approach described above. The same problem applies of optimal partitioning, and mitigating for sub-integrations with low signal strength.

\subsection{Combining single-pulse localisations}
If multiple consecutive pulses from a transient or pulsar are detected in multiple beams, log-likelihood maps for individual pulses may be added together to produce a more precise aggregate localisation. Our tests of this approach (see Bezuidenhout, in prep.) have shown that combining many single-pulse localisations, each with a relatively large uncertainties but similar means, may produce a better result than any individual pulse. Conversely, one high-S/N pulse’s localisation may only be deteriorated by combining it with those of lower-S/N detections. This effect is partially mitigated by weighting the individual localisations by S/N, but the method's utility remains to be judged on a case-by-case basis.

\subsection{Incorporating non-detections in adjacent TABs}
Real-time surveys like MeerTRAP often do not record S/N values in TABs where the S/N does not meet a certain threshold value. In those circumstances, it may be desirable to use the threshold S/N as an upper limit value in those TABs. Non-detections may then be used to further constrain the source position, especially if any non-detection TAB PSF has sensitive side-lobes coincident with a detection TAB. However, the MLE approach followed in this work does not allow for upper limits to be used.

Additionally, for non-detections to be successfully included in the localisation, it will require that the single-pulse detection is complete above the specified S/N threshold. We caution that incorporating a false negative in the single-pulse detection process, for whatever reason, would be detrimental to the localisation.

\subsection{Spectral index localisation}\label{sec:spec_ind}

\begin{figure*}
    \centering
    \includegraphics[width=\textwidth]{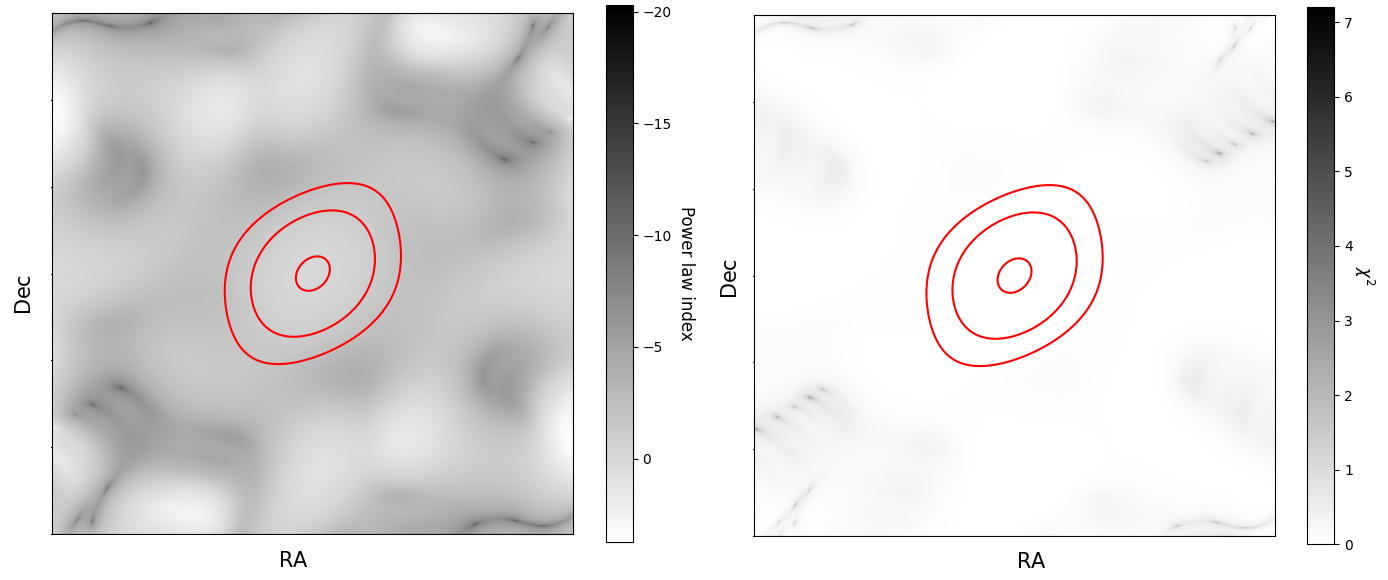}
    \caption{Spectral indices (left) and $\chi^{2}$ values (right) for power-law fits to the TAB PSF displayed in Fig.~\ref{fig:tilesim} using eight sub-bands. The red contours indicate the 25~per~cent, 50~per~cent, and 95~per~cent levels of the PSF at 1.284~GHz.}
    \label{fig:spec_maps}
\end{figure*}

\citet[][]{2015obrockaThesis} proposes that in order to constrain source positions, in addition to using the TAB PSFs as explained in $\S\ref{sec:tablo}$, the source's apparent spectral index as a function of RA and Dec may be used for the same purpose. In the formulation of \citet[][]{2015obrockaThesis}, the observed single-pulse flux density in a given TAB can be assumed to be described by a power law equation
\begin{equation}
    S(\nu) \propto  \nu^{\alpha},
\end{equation}\label{eq:alpha}
where $\nu$ is the frequency and $\alpha$ is the spectral index. The observed spectrum is assumed to be the product of two other power law spectra, namely that intrinsic to the source, with index $\alpha_{\mathrm{I}}$, and that imparted by the frequency dependence of the TAB PSF\footnote{Technically, the response of the receiver imparts a third spectral component, but this is inseparable from the TAB response without calibration, and, assuming it is constant across TABs, drops out when taking the index difference. Hence we ignore this component here.}, with index $\alpha_{\mathrm{TAB}}$, such that
\begin{equation}
    \alpha = \alpha_{\mathrm{I}} + \alpha_{\mathrm{TAB}}. 
\end{equation}\label{eq:alpha_sum}
Since the difference between spectral indices measured in two TABs is therefore independent of the intrinsic spectrum, the TABLo method described in $\S\ref{sec:tablo}$ can easily be adapted to fit for $\alpha$ differences rather than ratios of S/N values. The method described in $\S\ref{sec:tablo}$ can then be applied without any further changes. The resulting spectral index localisation can either be combined with the S/N localisation or used as a separate point of reference. The left-hand side panel of Fig.~\ref{fig:spec_maps} shows an example spectral index map for the PSF shown in Fig.~\ref{fig:tilesim}. 

It is important to note, however, that this localisation method rests on the assumption that the TAB sensitivity spectrum is well-fit by a power-law at all positions. This assumption may not hold, particularly away from the TAB centre beyond the main lobe, where the positions of sidelobes are strongly frequency dependent. This effect can be seen in the right-hand side panel of Fig.~\ref{fig:spec_maps}, which shows the $\chi^{2}$ values for the example power law fit. Note also that the strength of a spectral power law fit will depend strongly on the array configuration---arrays with non-Gaussian uv coverage will result in a complex TAB with plateaus, causing the spectrum to diverge from a power law.

Our tests (see Bezuidenhout; in prep) have shown that power-law index fitting can provide a good localisation of comparable precision to S/N fitting in the case that only closely-overlapping TABs with centres nearby the source position are used. However, for source positions far away from the centre of an included TAB, the localisation accuracy can become very poor. For spectral indices to be successfully incorporated into the localisation procedure, more complicated spectral fitting would be required, and the accompanying errors would need to be accounted for.

%-----------------------------------------------------------SECTION 6---------------------------------------------------------------------------------------------------------------

\section{Discussion and Conclusions}
% We have described a novel transient localisation method using the PSF of a TAB. We have also provided examples of its successful use in various idealised and real-world cases. Finally, we have outlined various ways in which this method may be elaborated and improved.

We have described TABLo, a novel transient localisation method facilitated by the modelling of interferometer TAB PSFs. $\S\ref{subsec:results_ideal}$ illustrates the performance of TABLo under both favourable and unfavourable circumstances. In cases of bright detections in many TABs at once, the derived likelihood distributions coincide closely with what is expected assuming 1-$\sigma$ Gaussian variance of the observed S/N values. If the pulse was weakly detected in the minimum of three TABs, however, we see some significant departures from Gaussianity. The positional likelihood distributions are then noticeably skewed, and the errors are not very well described by the standard deviations predicted by \textsc{SeeKAT}. Hence, in the event that a pulse is weakly detected in a few TABs, and especially if the predicted likelihood distribution is visibly asymmetrical, we recommend that users report the 2-$\sigma$ uncertainty on \textsc{SeeKAT} localisations rather than the 1-$\sigma$ uncertainty.

% The example idealised localisations and Gaussian random perturbation analysis presented in $\S\ref{subsec:results_ideal}$ illustrate the performance of TABLo under both favourable and unfavourable circumstances. In the cases where the TABs are very tightly packed, resulting in bright detections in many TABs at once, the \textsc{SeeKAT} likelihood distributions coincide closely with the expected distributions based on the Gaussian perturbation test. In the case that the source position is somewhat offset from the centre of the tiling pattern, the distribution of offsets is slightly skewed in one direction, although the 1-$\sigma$ and 2-$\sigma$ uncertainties are still close to the expected values.

% The cases where the pulse was weakly detected in only three TABs, however, show some significant departures from the assumption of Gaussianity that the uncertainty estimation rests on. For both the 25~per~cent and 50~per~cent localisations the offset distributions were noticeably skewed, and their standard deviations did not closely match those predicted by \textsc{SeeKAT}. Hence, we recommend that, in the event that a pulse is weakly detected in a small number of TABs, and especially if the predicted likelihood distribution is visibly asymmetrical, users report the 2-$\sigma$ uncertainty on \textsc{SeeKAT} localisations rather than the 1-$\sigma$ uncertainty.

We have also presented a number of real-world uses of \textsc{SeeKAT} for which the results could be compared to localisations using other methods. It is encouraging that the \textsc{SeeKAT} localisations agreed with the independent positions to within at least the 3-$\sigma$ uncertainty level. These represent the most precise non-image plane interferometric localisations of transient single pulses that have yet been achieved, sufficient for rapid follow-up observations. The utility of TABLo is further illustrated by the use of \textsc{SeeKAT} to phase-connect the timing solution of a new GC pulsar, as described in \citet[][]{2022vleeschower}.

% We also made various suggestions for future work on TABLo, which by including more pulse information can improve the localisation. A problem with such elaborations is maintaining the assumption of independent, normally distributed S/N values that the likelihood estimation introduced in this work rests on. Following different approaches to TABLo that do not make those assumptions, such as Bayesian MCMC inference, may be more fruitful in utilising the full range of information available for single-pulse transient and pulsar detections towards localisation. 

Finally, we will list the major factors that will determine the efficacy of \textsc{SeeKAT} for the consideration of prospective users:
\begin{enumerate}
    \item The number of beams in which the pulse is detected. This is affected by the intrinsic brightness of the pulse, the position of the source relative to the TAB tiling, at what level of response the TABs intersect, and the chosen threshold for considering a detection real. We would recommend that projects who aim to make use of this or an analogous method for localisation take this factor into account when deciding on their TAB tiling and single-pulse detection strategies.
    \item The accuracy of the S/N measurements. These are inherently imprecise, and may differ significantly depending on how the data are processed, the method of RFI excision, and how the value is determined. However, imprecisions should be mostly consistent across TABs, so that they are at least partially taken care of by dealing with the ratios of the values. Additionally, the higher the S/N value in a given beam the smaller the effect of any imprecision would be, so this method is certainly best suited for brighter pulses. It should be noted that instead of S/N values per se, theoretically any value proportional to the brightness of a source in a beam could be used to equal effect. If the noise factor differs significantly from TAB to TAB, then S/N may be too volatile a measurement, and another quantity proportional to the signal strength may give better results.
    \item The accuracy of the beamforming simulation. Since the S/N is averaged over the full band or a subsection of the band, while the PSF is generated for a single frequency, there is an inherent error in comparing the ratios of PSFs to the ratios of S/N values. This error will be larger in the case of a sidelobe detection, since the sidelobes are much more frequency dependent than the TAB's main lobe. This effect is partially mitigated in \textsc{SeeKAT} by setting values of the PSF below a certain value (by default 8~per~cent of the maximum) to zero. We also recommend that users discard parts of the band where the pulse is not visible before the S/N calculation, as well as to generate the PSF for where the pulse is brightest rather than necessarily at the centre of the band. It is important to note that the error from this imprecision, as well as any other possible error on the PSF model, is not included in the \textsc{SeeKAT} uncertainty calculation. 
\end{enumerate}

We have illustrated the value of beamforming simulations for localising sources without the need for imaging. This is sure to prove useful considering the cost of storing large amounts of baseband data. We have also developed a suite of software\footnote{\url{https://github.com/BezuidenhoutMC/SeeKAT}} for accomplishing this task, which has already been used by other MeerKAT projects for localising newly-discovered sources. This work should be easily adaptable to other interferometers, and could see increasing use in the coming SKA-era of radio astronomy. 

\section*{Acknowledgements}

M.C.B., B.W.S., F.J., K.R., and M.S. thank the MeerKAT LSP teams for allowing MeerTRAP to observe commensally. The MeerTRAP project has received funding from the European Research Council (ERC) under the European
Union’s Horizon 2020 research and innovation programme (grant
agreement No. 694745). KMR acknowledges support from the Vici research program 'ARGO' with project number 639.043.815, financed by the Dutch Research Council (NWO). C.J.C. and R.P.B. acknowledge support from the ERC under the European Union’s Horizon 2020 research and innovation programme (grant agreement No. 715051; Spiders). The MeerKAT telescope is operated by the
South African Radio Astronomy Observatory, which is a facility
of the National Research Foundation, an agency of the Department
of Science and Innovation.

%%%%%%%%%%%%%%%%%%%%%%%%%%%%%%%%%%%%%%%%%%%%%%%%%%
\section*{Data Availability}

The data will be made available to others upon reasonable request
to the authors.



\bibliographystyle{rasti}
\bibliography{main} % if your bibtex file is called example.bib

%\begin{thebibliography}{99}
%\end{thebibliography}

%%%%%%%%%%%%%%%%%%%%%%%%%%%%%%%%%%%%%%%%%%%%%%%%%%

%%%%%%%%%%%%%%%%% APPENDICES %%%%%%%%%%%%%%%%%%%%%

% \appendix

% \section{Some extra material}

% If you want to present additional material which would interrupt the flow of the main paper,
% it can be placed in an Appendix which appears after the list of references.

%%%%%%%%%%%%%%%%%%%%%%%%%%%%%%%%%%%%%%%%%%%%%%%%%%

% Don't change these lines
\bsp	% typesetting comment
\label{lastpage}
\end{document}